\documentclass[12pt, draftclsnofoot, onecolumn, twoside]{IEEEtran}
\usepackage{amssymb,amsmath,mathrsfs,bm,graphicx,color,cite,subfigure,multirow,cases}
\usepackage{hyperref}

\allowdisplaybreaks

\newtheorem{lemma}{Lemma}
\newtheorem{theorem}{Theorem}

\newtheorem{remark}{Remark}

\newcommand{\by}{ {\bf y} }
\newcommand{\bu}{ {\bf u} }
\newcommand{\bv}{ {\bf v} }

\newcommand{\bi}{ {\bf i} }
\newcommand{\bn}{ {\bf n} }

\newcommand{\pardef}{\stackrel{\Delta}{=}}
\newcommand{\e}{ \mathrm{e} }
\newcommand{\diff}{ \mathrm{d} }

\begin{document}
\title{Noncoherent Relaying in Energy Harvesting Communication Systems}
\author{Peng Liu,  
        Saeed Gazor, 
        Il-Min Kim, 
        and Dong In Kim
%
\thanks{P. Liu is is   with   the   Department   of   Electrical   Engineering,   Stanford   University,   Stanford,   CA,   94305,   USA   (e-mail:
pengliu1@stanford.edu).}
\thanks{I.-M. Kim and S.Gazor are with the Department of Electrical and Computer Engineering, Queen's University, Kingston, Ontario, K7L 3N6, Canada (e-mail: ilmin.kim@queensu.ca; gazor@queensu.ca)}
\thanks{D. I. Kim is with the school of Information and Communication Engineering, Sungkyunkwan University (SKKU), Suwon, Korea (e-mail: dikim@skku.ac.kr)}
}

\markboth{IEEE Transactions on Wireless Communications}
{Liu \MakeLowercase{\emph{et al.}}: Noncoherent Relaying in Energy Harvesting Communication Systems}


\maketitle

\begin{abstract}
In energy harvesting (EH) relay networks, the \emph{coherent} communication requires accurate estimation/tracking of the \emph{instantaneous} channel state information (CSI) which consumes extra power. As a remedy, we propose two \emph{noncoherent} EH relaying protocols based on the amplify-and-forward (AF) relaying, namely, power splitting noncoherent AF (PS-NcAF) and time switching noncoherent AF (TS-NcAF), which do not require any instantaneous CSI. We develop a \emph{noncoherent} framework of simultaneous wireless information and power transfer (SWIPT), embracing PS-NcAF and TS-NcAF in a \emph{unified} form. For arbitrary $M$-ary noncoherent frequency-shift keying (FSK) and differential phase-shift keying (DPSK), we derive maximum-likelihood detectors (MLDs) for PS-NcAF and TS-NcAF in a \emph{unified} form, which involves integral evaluations yet serves as the optimum performance benchmark. To avoid expensive integral computations, we propose a \emph{closed-form} detector using the Gauss-Legendre approximation, which achieves almost identical performance as the MLD but at substantially lower complexity.  These EH-based noncoherent detectors achieve full diversity  in Rayleigh fading. Numerical results demonstrate that our proposed PS-NcAF and TS-NcAF may outperform the conventional grid-powered relay system under the same total power constraint. Various insights which are useful for the design of practical SWIPT relaying systems are obtained. Interestingly, PS-NcAF outperforms TS-NcAF in the single-relay case, whereas TS-NcAF outperforms PS-NcAF in the multi-relay case.
\end{abstract}


 \newpage

\section{Introduction}  \label{sec:intro}
Conventional battery powered wireless communications systems require periodic recharging or replacement of the batteries, which incurs a high operation burden \cite{Q.Dong2015.2.accepted} and can be cumbersome or even impossible (e.g., for biomedical devices implanted in the human body \cite{S.Kim2012.8}). Recently, energy harvesting (EH) from the ambient radio-frequency (RF) signals has been developed as one of the attractive alternatives to prolong the lifetime of energy-constrained nodes in wireless networks \cite{N.Shinohara2014book, H.J.Visser2013.6}. The dual use of RF signals for EH and information delivery (ID) has led to the novel architecture of simultaneous wireless information and power transfer (SWIPT) \cite{L.R.Varshney2008.7,P.Grover2010.6}, which allows wireless nodes to scavenge energy as well as extract information simultaneously from the RF signals, thus constituting an appealing solution for energy-constrained applications such as wireless relay networks.

Practical SWIPT receiver architectures make use of two different circuits performing EH and ID individually \cite{R.Zhang2013.5}. The receiver may either switch between the EH and ID circuits in a time-division fashion, a scheme known as time switching \cite{L.Liu2013.1}, or split the received RF signals into two streams fed to the EH and ID circuits at the same time, a scheme known as power splitting \cite{L.Liu2013.9}.  Typically,  EH and ID circuits operate with rather different receiver power sensitivities, i.e., $-60$ dBm for information receivers and $-10$ dBm for energy receivers  \cite{X.Lu.accepted.Kim,A.A.Nasir.submitted}. 
Since the EH circuit is designed to maximize the EH efficiency while the ID circuit typically aims for maximum information rate, a fundamental \emph{rate-energy tradeoff} exists for SWIPT systems \cite{L.Liu2013.1,X.Zhou2013.11, L.Liu2013.9,R.Zhang2013.5}. Various resource allocation and beamforming schemes were designed to achieve different tradeoffs between rate and energy in SWIPT systems. In particular, the optimum power allocation maximizing the information rate subject to EH constraint was studied for broadband systems with perfect instantaneous channel state information (CSI) \cite{K.Huang2013.12}. The optimum resource allocation with service differentiation between low- and high-priority data was studied for EH networks in \cite{X.Lu.accepted}. In \cite{D.Niyato2014.9.accepted}, a joint optimization and game-theoretic framework was developed to optimize the packet delivery policy and cooperation strategy for delay tolerant networks with EH. Moreover, beamforming schemes which maximize the information rate subject to EH constraint or maximize the harvested energy subject to rate constraint were studied under imperfect CSI in \cite{C.Xing2013.4} and \cite{Z.Xiang2012.8}, respectively.


Recently, SWIPT has also been applied in wireless relay systems which allows energy-constrained relay nodes to harvest energy from the source RF signals and consequently brings in substantial benefits in wireless networks \cite{B.Medepally2010.11}. Various theoretical analysis and practical design have been conducted for EH relay systems in the literature. Specifically, for dual-hop amplify-and-forward (AF) networks, a greedy switching protocol between data relaying and EH was proposed in \cite{I.Krikidis2012.11}, and the joint relay selection and power allocation scheme was developed in \cite{I.Ahmed2013.4}. Moreover, the outage probability and ergodic capacity for EH AF relaying were derived for delay-limited and delay-tolerant applications, respectively, in \cite{A.A.Nasir2013.7}. Besides AF relay systems, the EH has also been considered for decode-and-forward (DF) networks. In particular, the outage probability of EH DF relaying was studied, taking into account the spatial randomness of the source-destination pairs, for various network topologies in \cite{Z.Ding2013.12, Z.Ding2014.2, Z.Ding2014.8, I.Krikidis2014.3}.

The aforementioned studies on EH relay systems have either implicitly or explicitly assumed the instantaneous CSI availability to allow for \emph{coherent} information decoding. However, the coherent SWIPT requires the source to periodically send training symbols, which incurs an increased signaling overhead and processing burden. Moreover, the relays may need to relay the training symbols and/or estimate the source-relay channels \cite{P.Liu2010.1}, which results in additional energy consumptions and poses serious issues especially for energy-constrained relay nodes. As a remedy, the \emph{noncoherent} SWIPT eliminating the need for the instantaneous CSI was first studied for the EH DF relay systems \cite{P.Liu2014.12.submitted}, where two EH relaying protocols, namely, the power splitting noncoherent DF (PS-NcDF) and time switching noncoherent DF (TS-NcDF) were proposed, and the corresponding maximum-likelihood detectors (MLDs) facilitating noncoherent SWIPT were obtained. However, these protocols and the detectors are applicable only for the DF relay systems. It is still unknown about how the noncoherent AF relaying can benefit from EH, and how the noncoherent AF relaying performs as compared to the noncoherent DF relaying in EH relay systems. In addition, for EH AF relay systems, the noncoherent MLD minimizing the symbol-error rate (SER) in noncoherent SWIPT is still unknown. Realizing that none of the previous works have tackled the design challenges of noncoherent SWIPT in EH AF relay systems, we aim to fill the gap.


In this paper, we study the \emph{noncoherent} EH relay systems consisting of multiple AF relays, which can harvest energy from the ambient source signals and utilize the harvested energy to assist the communication. The main contributions of this paper are summarized as follows:

\begin{itemize}
  \item For arbitrary $M$-ary noncoherent frequency-shift keying (FSK) or differential phase-shift keying (DPSK) signalings, we propose a unified \emph{noncoherent} SWIPT framework embracing two EH relaying protocols, namely, the power splitting noncoherent AF (PS-NcAF) and time switching noncoherent AF (TS-NcAF).
  \item Following the proposed SWIPT framework, we develop noncoherent MLDs for PS-NcAF and TS-NcAF in a \emph{unified} form, which involves integral evaluations yet serves as the optimum performance benchmark for noncoherent SWIPT. To avoid integral computations, we also develop a \emph{closed-form} Gauss-Legendre approximation based detector, which achieves almost identical SER to the MLD at substantially lower complexity. It is demonstrated that the proposed EH-based noncoherent detectors achieve full diversity  in Rayleigh fading. In terms of the error performance, the proposed PS-NcAF and TS-NcAF may outperform conventional grid-powered noncoherent relay systems under the same total power constraint. 
  \item The choice of the time switching or power splitting parameters represents a tradeoff between EH and information transmission, and there exist unique optimum values of these parameters which minimize the SER.
  \item The number of EH relay nodes is a key factor on the performance of the noncoherent EH relay systems: PS-NcAF outperforms TS-NcAF in the single-relay case, whereas TS-NcAF outperforms PS-NcAF in the multi-relay case.
  \item The optimum relay position is closer to the source than to the destination, regardless of the EH relaying protocols (This is in contrast to the conventional self-powered relay systems where the optimum relay position is closer to the destination for AF and is closer to the source for DF). Furthermore, as the path-loss exponent increases, the optimum relay position shifts slightly towards the destination, but is still closer to the source.
  \item The $M$-FSK signaling with $M\geq 8$ is a more suitable solution for EH relay systems due to its higher energy efficiency compared to $M$-DPSK, regardless of the EH relaying protocols.
  \item Comparison of the AF- and DF-based EH relaying protocols demonstrates that the SER performance of AF and DF relaying is almost the same in noncoherent EH relay systems, irrespective of the SWIPT architecture (e.g., power splitting or time switching). This is in contrast to the conventional self-powered relay systems where either AF or DF may outperform each other depending on the relay locations.
\end{itemize}

The remainder of the paper is organized as follows. Section \ref{sec:sysmodel} describes the system model and develops noncoherent EH relaying protocols. Section \ref{sec:unifiedSysModel} obtains a unified noncoherent SWIPT framework for EH AF relay system. Section \ref{sec:MLD} derives the noncoherent detectors achieving SWIPT. Section \ref{sec:numerical} gains useful insights into the noncoherent SWIPT through simulations and Section \ref{sec:con} concludes the paper.

\emph{Notation}: We use $(\cdot)^*$, $(\cdot)^T$, $(\cdot)^H$,  $\Re(\cdot)$, $\mathbb{E}(\cdot)$, $\|\cdot\|$, and $\ln(\cdot)$ to denote the conjugate, transpose, conjugate transpose, expectation, real part, 2-norm, and natural logarithm, respectively. Also, $\bm{0}$, $\bm{I}_n$, and $\bi_{n}$ denote an all-zero column vector, an $n\times n$ identity matrix, and a column vector with 1 at its $n$-th entry and 0 elsewhere, respectively. Moreover, $A \pardef B$ denotes that $A$ is defined by $B$. Finally, $\bm{x}\sim\mathcal{CN}(\bm{\mu},\bm{\Sigma})$ means that $\bm{x}$ is a circularly symmetric complex Gaussian (CSCG) random vector with mean $\bm{\mu}$ and covariance $\bm{\Sigma}$.


\section{System Description and Noncoherent EH Relaying Protocols}\label{sec:sysmodel}
Consider an EH relay network where the source terminal $\mathrm{T}_0$ communicates with the destination terminal $\mathrm{T}_d$ through the help of a set of potential relay candidates, where the selected relays can harvest energy from the source RF signals and forward the source information. Suppose that a total of $K$ EH relays, $\mathrm{T}_r$, $r=1,2,\cdots,K$, are predetermined.\footnote{Any potential relay selection protocols may be adopted. For example, a natural method is to choose the relays whose harvested energy is above certain threshold. Note that our proposed EH relaying protocols are valid for any specific relay selection schemes, and a detailed treatment of the relay selection scheme is beyond the scope of this paper.} The source may communicate directly with the destination through the source-destination link $h_{0d}$ or indirectly via the two-hop relay channels $h_{0r}$ (first hop) and $h_{rd}$ (second hop). The relays, which do not have fixed power supplies, can only harvest energy from the RF signals radiated by the source and utilize that harvested energy to assist the source-destination communication through the noncoherent AF relaying, thus enabling SWIPT in the EH relay networks. We consider a composite fading model comprising the large-scale path loss $\mathcal{L}_{ij}$ as well as the small-scale fading $h_{ij}$, $ij\in\{0d,0r,rd\}_{r=1}^K$. The path loss component $\mathcal{L}_{ij}$ is a distance-dependent constant and will be specified later. The small-scale fading coefficients are modeled as $h_{ij}\sim\mathcal{CN}(0,1)$, which corresponds to the Rayleigh fading scenario. We assume that the instantaneous CSIs, $h_{ij}$, $ij\in\{0d,0r,rd\}_{r=1}^K$, are unavailable to any terminals in the network, and thus, \emph{noncoherent} signalings such as $M$-FSK and $M$-DPSK are considered. Note that the quadrature amplitude modulation (QAM) is inconsistent with the noncoherent SWIPT framework considered here as it typically requires the instantaneous CSI at the receiver for coherent information delivery.

\begin{figure*}[!t]%
\centering
\subfigure[][]{\includegraphics[width=0.5\columnwidth]{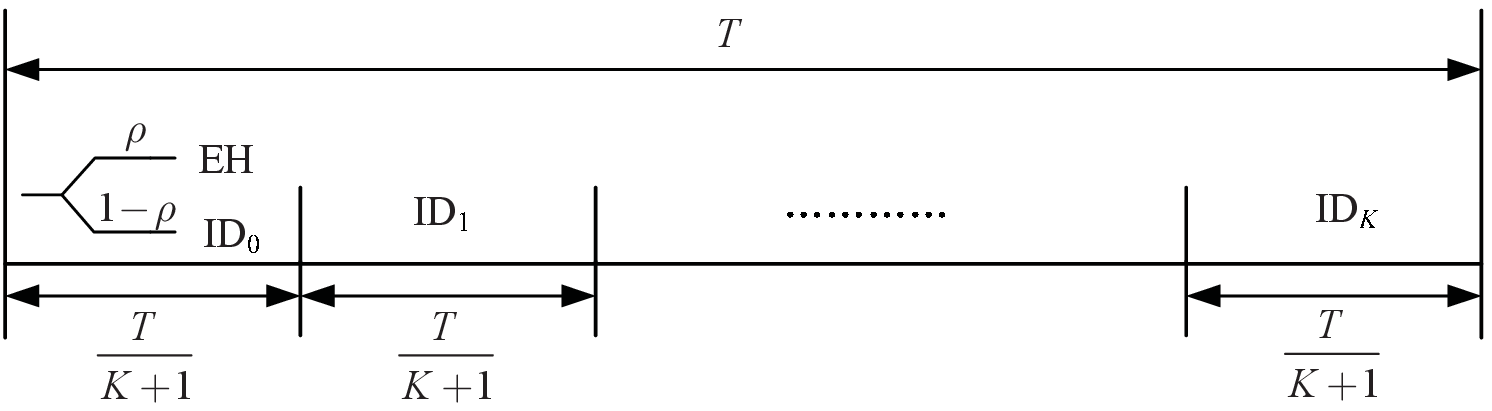}\label{fig:PS_NcAF}} \hfil
\subfigure[][]{\includegraphics[width=0.5\columnwidth]{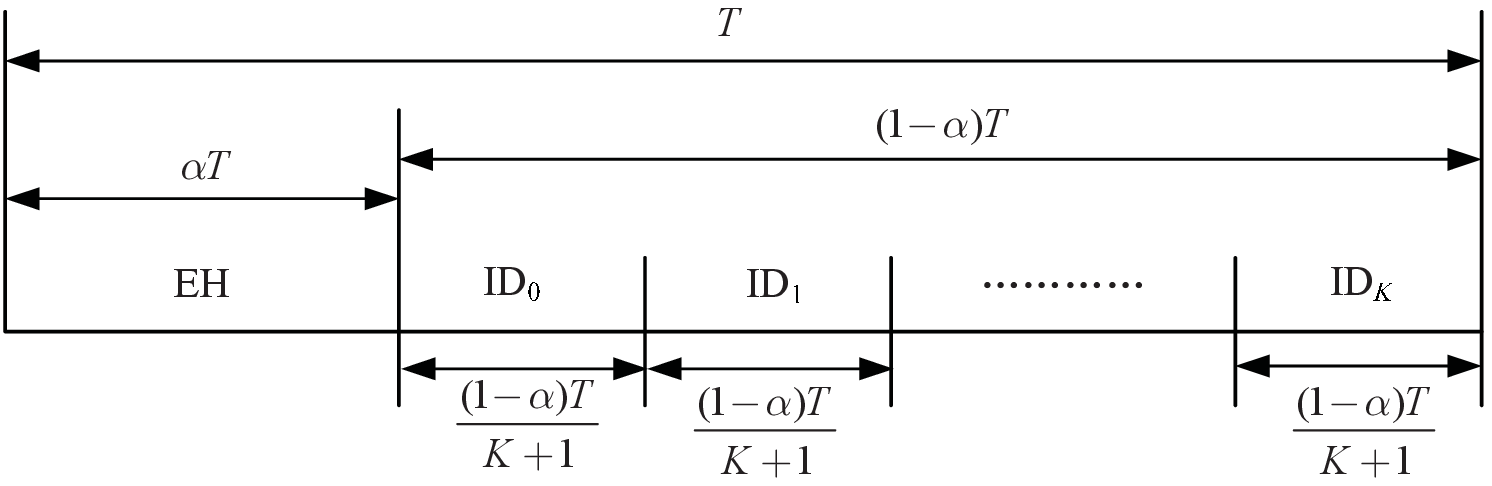}\label{fig:TS_NcAF}}
\caption{EH $K$-relay systems with simultaneous wireless information and power transfer, where ID$_k$ denotes the information delivery (or data relaying) from $\mathrm{T}_k$, $k=0,\cdots.K$. \subref{fig:PS_NcAF} PS-NcAF where each relay splits its received RF power into two portions: the $\rho$ portion is for EH and the remaining $1-\rho$ portion is for ID. \subref{fig:TS_NcAF} TS-NcAF where the total block time is divided into the EH phase of length $\alpha T$ and the ID phase of length $(1-\alpha)T$.}.
\label{fig:sysmodel}
\end{figure*}

\subsection{PS-NcAF Relaying Protocol}\label{sub:PS_NcAF}
In the PS-NcAF protocol, each EH relay node splits its received RF signal into two streams, which are fed to the EH and ID circuits at the same time. Suppose that the total communication block time $T$ (sec) is divided into $K+1$ sub-blocks for PS-NcAF, each of length $\frac{T}{K+1}$ is allocated to one of the transmitters (including the source and relays), as illustrated in Fig. \ref{fig:PS_NcAF}. In the first sub-block, the source broadcasts the RF signal with power $P_0$ (Watts). At each EH relay node, the $\rho$ portion of the received RF signal is used for EH, and the harvested energy is used for relaying the source signal in one of the following $K$ sub-blocks; thus, the average harvested power $P_r$ available for data relaying at $\mathrm{T}_r$ is 
\begin{align}
  P_r=\eta \rho P_0\mathcal{L}_{0r},\label{eq:Pk_PS}
\end{align}
 $r=1,\cdots,K$, where $0<\rho<1$ is the power splitting factor (PSF) and $0<\eta<1$ is the EH efficiency \cite{X.Zhou2013.11}. The remaining $1-\rho$ portion of the received RF signal is fed to the ID circuit at each relay node, where the signal is amplified using the harvested power $P_r$ in \eqref{eq:Pk_PS} and forwarded to the destination in one of the following $K$ sub-blocks.

\subsubsection{PS-NcAF with $M$-DPSK}
For $M$-DPSK transmission, each source message $m\in \{0,\cdots,M-1\}$ is differentially encoded into two consecutive information-bearing symbols $s(l)$ and $s(l-1)$ according to $s(l)=s(l-1)\e^{j2\pi m/M}$, where $s(0)=1$ is the initial reference signal. Since the detection of each source message $m$ is based on two consecutive received symbols at the destination, it is convenient to represent the signals as $2\times 1$ vectors. Let $\bm{s}\pardef [s(l-1),s(l)]^T$ denote the source transmitted signal. Then, the received signal, $\bm{y}_{0r}\pardef [y_{0r}(l-1),y_{0r}(l)]^T$, at the relay node $\mathrm{T}_r$, $r=1,\cdots,K$, which is the intended for ID, can be expressed as 
\begin{align}
  \bm{y}_{0r}=\sqrt{(1-\rho)P_0T_s\mathcal{L}_{0r}}h_{0r}\bm{s}+\sqrt{1-\rho}\bm{u}_{0r}+\bm{v}_{0r},\label{eq:y0r_DPSK}
\end{align}
where $T_s$ (sec) is the symbol time and the power-scaling factor $1-\rho$ is due to the power splitting at $\mathrm{T}_r$. The additive white Gaussian noise (AWGN) at $\mathrm{T}_r$ (when $\mathrm{T}_0$ serves as the transmitter) is made up of two components: the AWGN due to the receive antenna (which is introduced before the power splitter), modeled as $\bm{u}_{0r}\sim\mathcal{CN}(0,\sigma_{0r,1}^2\bm{I}_2)$, and the AWGN due to the ID circuit (which is introduced after the power splitter), modeled as $\bm{v}_{0r}\sim\mathcal{CN}(0,\sigma_{0r,2}^2\bm{I}_2)$ \cite{L.Liu2013.9,K.Huang2013.12,X.Zhou2013.11}. At the destination, no power splitting is needed as the received signal is only used for ID. Thus, the received sinal at the destination can be written as 
\begin{align}
  \bm{y}_{0d}=\sqrt{P_0T_s\mathcal{L}_{0d}}h_{0d}\bm{s}+\bm{u}_{0d}+\bm{v}_{0d},\label{eq:y0d_DPSK}
\end{align}
where $\bm{u}_{0d}\sim\mathcal{CN}(0,\sigma_{0d,1}^2\bm{I}_2)$ and $\bm{v}_{0d}\sim\mathcal{CN}(0,\sigma_{0d,2}^2\bm{I}_2)$ are the AWGNs due to the receive antenna and ID circuit, respectively.

Each relay amplifies its received signal $\bm{y}_{0r}$ with an amplifying gain $G_r$ and forwards that signal to the destination. Thus, the received signal at the destination is given by
\begin{align}
  \bm{y}_{rd}=\sqrt{(1-\rho)P_0T_s\mathcal{L}_{0r}\mathcal{L}_{rd}}G_rh_{0r}h_{rd}\bm{s}+G_r\sqrt{\mathcal{L}_{rd}}h_{rd}\big(\sqrt{1-\rho}\bm{u}_{0r}+\bm{v}_{0r}\big)
  +\bm{u}_{rd}+\bm{v}_{rd},\label{eq:yrd_DPSK}
\end{align}
where $\bm{u}_{rd}\sim\mathcal{CN}(0,\sigma_{rd,1}^2\bm{I}_2)$ and $\bm{v}_{rd}\sim\mathcal{CN}(0,\sigma_{rd,2}^2\bm{I}_2)$ are the AWGNs due to the receive antenna and ID circuit, respectively.
The amplifying gain $G_r$ ensures that the average transmission power for data relaying at $\mathrm{T}_r$ is fixed to $P_r$ in \eqref{eq:Pk_PS}  \cite{P.Liu2013.9}. For PS-NcAF with $M$-DPSK, the amplifying gain $G_r$ is given by
\begin{align}
  G_r=\sqrt{\frac{P_rT_s}{\mathbb{E}\{|y_{0r}(l)|^2\}}}=\sqrt{\frac{\eta\rho P_0T_s\mathcal{L}_{0r}}{(1-\rho)P_0T_s\mathcal{L}_{0r}+(1-\rho)\sigma_{0r,1}^2+\sigma_{0r,2}^2}}.\label{eq:PS_AF_DPSK_gain}
\end{align}

\subsubsection{PS-NcAF with $M$-FSK}
For $M$-FSK transmission, the message $m\in \{0,\cdots,M-1\}$ is transmitted over one of the $M$ orthogonal carriers. The baseband equivalent received signal intended for ID at the receiving terminal $\mathrm{T}_j$ which is transmitted from $\mathrm{T}_i$ is denoted as an $M\times 1$ vector $\by_{ij}\pardef[\mathrm{y}_{ij}(1),\cdots,\mathrm{y}_{ij}(M)]^T$, $ij\in\{0d,0r,rd\}_{r=1}^N$. Then, the signal model for PS-NcAF employing noncoherent $M$-FSK can be represented as
\begin{subequations}
  \begin{align}
  \by_{0d}&=\sqrt{P_0T_s\mathcal{L}_{0d}}h_{0d}\bi_{m+1}+\bu_{0d}+\bv_{0d},\label{eq:by_0d} \\
  \by_{0r}&=\sqrt{(1-\rho)P_0T_s\mathcal{L}_{0r}}h_{0r}\bi_{m+1}+\sqrt{1-\rho}\bu_{0r}+\bv_{0r},\label{eq:by_0r}\\
  \by_{rd}&=\sqrt{(1-\rho)P_0T_s\mathcal{L}_{0r}\mathcal{L}_{rd}}G_rh_{0r}h_{rd}\bi_{m+1}+G_r\sqrt{\mathcal{L}_{rd}}h_{rd}\big(\sqrt{1-\rho}\bu_{0r}+\bv_{0r}\big)
  +\bu_{rd}+\bv_{rd},\label{eq:by_rd}
\end{align}
\end{subequations}
where $\bu_{ij}\sim\mathcal{CN}(\bm{0},\sigma_{ij,1}^2\bm{I}_M)$ and $\bv_{ij}\sim\mathcal{CN}(\bm{0},\sigma_{ij,2}^2\bm{I}_M)$, $ij\in\{0d,0r,rd\}_{r=1}^N$, are the AWGNs due to the receive antenna and the ID circuit at $\mathrm{T}_j$, respectively. For PS-NcAF with $M$-FSK, the amplifying gain $G_r$, which ensures that the average transmission power for data relaying at $\mathrm{T}_r$ is fixed to $P_r$ in \eqref{eq:Pk_PS}, is given by \cite{P.Liu2013.9}
\begin{align}
  G_r= \sqrt{\frac{P_rT_s}{\mathbb{E}\{\|\by_{0r}\|^2\}}} =\sqrt{\frac{\eta\rho P_0T_s\mathcal{L}_{0r}}{(1-\rho)P_0T_s\mathcal{L}_{0r}+M\big[(1-\rho)\sigma_{0r,1}^2+\sigma_{0r,2}^2\big]}}.\label{eq:PS_AF_FSK_gain}
\end{align}

\subsection{TS-NcAF Relaying Protocol}\label{sub:TS_NcAF}
In the TS-NcAF protocol, the total communication block time $T$ (sec) is divided into two consecutive phases: the EH phase of length $\alpha T$ and the ID phase of length $(1-\alpha)T$, as illustrated in Fig. \ref{fig:TS_NcAF}, where $0<\alpha<1$ is the time switching coefficient (TSC). In the EH phase, the source sends RF energy flow to the relays with power $P_0$. Each EH relay can harvest energy from the received RF signal and utilize that harvested energy for data relaying in the subsequent ID phase. The ID phase is further divided into $K+1$ sub-blocks of length $\frac{(1-\alpha)T}{K+1}$ each, where the first sub-block is assigned to the source for new information transmission and each of the remaining $K$ sub-blocks is assigned to one of the relays for data relaying. Then, the average harvested power $P_r$ for data relaying at $\mathrm{T}_r$ is given by
\begin{align}
  P_r=\frac{(K+1)\eta P_0\mathcal{L}_{0r}\alpha}{1-\alpha},\label{eq:Pk_TS}
\end{align}
$r=1,\cdots,K$, where $0<\eta\leq 1$ is the energy conversion efficiency \cite{X.Zhou2013.11}.


\subsubsection{TS-NcAF with $M$-DPSK}
In the TS-NcAF protocol, the received signal during the ID phase at each node is solely used for information processing, i.e., no power splitting is involved. Thus, the baseband equivalent signal model for TS-NcAF with $M$-DPSK is represented in ($2\times 1$)-vector form as follows:
\begin{subequations}
    \begin{align}
  \bm{y}_{0d}&=\sqrt{P_0T_s\mathcal{L}_{0d}}h_{0d}\bm{s}+\bm{u}_{0d}+\bm{v}_{0d},\label{eq:y0d_DPSK_TS}\\
  \bm{y}_{0r}&=\sqrt{P_0T_s\mathcal{L}_{0r}}h_{0r}\bm{s}+\bm{u}_{0r}+\bm{v}_{0r},\label{eq:y0r_DPSK_TS}\\
  \bm{y}_{rd}&=\sqrt{P_0T_s\mathcal{L}_{0r}\mathcal{L}_{rd}}G_rh_{0r}h_{rd}\bm{s}+G_r\sqrt{\mathcal{L}_{rd}}(\bm{u}_{0r}+\bm{v}_{0r})+\bm{u}_{rd}+\bm{v}_{rd}.
  \label{eq:yrd_DPSK_TS}
\end{align}
\end{subequations}
$r=1,\cdots,K$, where $\bm{u}_{ij}\sim\mathcal{CN}(0,\sigma_{ij,1}^2\bm{I}_2)$ and $\bm{v}_{ij}\sim\mathcal{CN}(0,\sigma_{ij,2}^2\bm{I}_2)$, $ij\in\{0d,0r,rd\}_{r=1}^K$, are the AWGNs due to the receive antenna and the ID circuit, respectively. For TS-NcAF with $M$-DPSK, the amplifying gain $G_r$, which ensures that the average transmission power at $\mathrm{T}_r$ is fixed to $P_r$ in \eqref{eq:Pk_TS}, is given by \cite{P.Liu2013.9}
\begin{align}
  G_r=\sqrt{\frac{(K+1)\eta P_0T_s\mathcal{L}_{0r}\alpha}{(1-\alpha)(P_0T_s\mathcal{L}_{0r}+\sigma_{0r,1}^2+\sigma_{0r,2}^2)}}.\label{eq:TS_AF_DPSK_gain}
\end{align}

\subsubsection{TS-NcAF with $M$-FSK}
The baseband equivalent signal model for TS-NcAF employing noncoherent $M$-FSK can be represented as
\begin{subequations}\label{eq:by0d_TS}
\begin{align}
  \by_{0d}&=\sqrt{P_0T_s\mathcal{L}_{0d}}h_{0d}\bi_{m+1}+\bu_{0d}+\bv_{0d},\label{eq:by_0d_TS}\\
  \by_{0r}&=\sqrt{P_0T_s\mathcal{L}_{0r}}h_{0r}\bi_{m+1}+\bu_{0r}+\bv_{0r},\label{eq:by_0r_TS}\\
  \by_{rd}&=\sqrt{P_0T_s\mathcal{L}_{0r}\mathcal{L}_{rd}}G_rh_{0r}h_{rd}\bi_{m+1}+G_r\sqrt{\mathcal{L}_{rd}}h_{rd}(\bu_{0r}+\bv_{0r})
  +\bu_{rd}+\bv_{rd},\label{eq:by_rd_TS}
\end{align}
\end{subequations}
for $r=1,\cdots,K$, where $\bu_{ij}\sim\mathcal{CN}(\bm{0},\sigma_{ij,1}^2\bm{I}_M)$ and $\bv_{ij}\sim\mathcal{CN}(\bm{0},\sigma_{ij,2}^2\bm{I}_M)$, $ij\in\{0d,0r,rd\}_{r=1}^K$, are the AWGNs due to the receive antenna and ID circuit at $\mathrm{T}_j$, respectively. For TS-NcAF with $M$-FSK, the amplifying gain $G_r$, which ensures that the average transmission power at $\mathrm{T}_r$ is fixed to $P_r$ in \eqref{eq:Pk_TS}, is given by \cite{P.Liu2013.9}
\begin{align}
  G_r=\sqrt{\frac{(K+1)\eta P_0T_s\mathcal{L}_{0r}\alpha}{(1-\alpha)\big[P_0T_s\mathcal{L}_{0r}+M(\sigma_{0r,1}^2+\sigma_{0r,2}^2)\big]}}.\label{eq:TS_AF_FSK_gain}
\end{align}

\section{A Unified Noncoherent SWIPT Framework}\label{sec:unifiedSysModel}
In the last section, we developed \emph{two} EH relaying protocols, namely, PS-NcAF and TS-NcAF, which can be easily applied in conjunction with \emph{two} widely adopted noncoherent signalings such as noncoherent $M$-FSK and $M$-DPSK, thus resulting in a total of \emph{four} different noncoherent SWIPT schemes for EH relay systems. From Section \ref{sec:sysmodel}, we see that each scheme has a different system model with a set of different parameters. In particular, the amplifying gain expressions $G_r$ corresponding to the four possible schemes in \eqref{eq:PS_AF_DPSK_gain}, \eqref{eq:PS_AF_FSK_gain}, \eqref{eq:TS_AF_DPSK_gain}, and \eqref{eq:TS_AF_FSK_gain} are all different. These distinct system models and inconsistent parameters make it rather cumbersome for further design and analysis. To resolve this problem, in this section, we will develop a unified SWIPT framework embracing both PS-NcAF and TS-NcAF, which enables unified further development.

The key to the unifying process is to unify the definitions of all different system parameters. First of all, we introduce the \emph{effective noise variance} for each receiving terminal as follows:
\begin{subequations}\label{eq:noise_variances}
\begin{align}
  \sigma_{0d}^2 &\pardef \sigma_{0d,1}^2 + \sigma_{0d,2}^2 \\
  \sigma_{rd}^2 &\pardef \sigma_{rd,1}^2 + \sigma_{rd,2}^2 \\
  \sigma_{0r}^2 & \pardef \left \{\begin{array}{ll}
                                    (1-\rho)\sigma_{0r,1}^2+\sigma_{0r,2}^2, & \text{PS-NcAF}, \\
                                    \sigma_{0r,1}^2+\sigma_{0r,2}^2, & \text{TS-NcAF}.
                                  \end{array}
   \right.
\end{align}
\end{subequations}
Furthermore, let us introduce the information rate, $R\pardef \frac{N_s\log_2 M}{T}$ (bps), which is the total transmitted information bits normalized by the total communication block time $T$ (sec), where $N_s$ is the total number of transmitted information-bearing symbols. For PS-NcAF, the $N_s$ new symbols are sent over the first sub-block of length $\frac{T}{K+1}$. For TS-NcAF, due to the time switching effect, the $N_s$ new symbols are transmitted over the first sub-block of length $\frac{(1-\alpha)T}{K+1}$. By definition of $R$, we have
\begin{align}
  R=\left\{
        \begin{array}{ll}
          \frac{\log_2M}{(K+1)T_s}, & \hbox{PS-NcAF}, \\
          \frac{(1-\alpha)\log_2M}{(K+1)T_s}, & \hbox{TS-NcAF}.
        \end{array}
      \right.\label{eq:rate}
\end{align}
where $T_s$ depends on $T$ through $N_s$ as follows:
\begin{align}
  T_s=\left\{\begin{array}{ll}
    \frac{T}{(K+1)N_s},& \text{PS-NcAF},\\
    \frac{(1-\alpha)T}{(K+1)N_s},& \text{TS-NcAF},
  \end{array}
  \right.\nonumber
\end{align}
For the purpose of performance comparison of PS-NcAF and TS-NcAF, one must ensure that the information rate $R$ is the same for both schemes, which is accomplished by choosing the common parameters $T$ (total communication block time), $N_s$ (total number of symbols), and $M$ (modulation alphabet size) for both schemes. This implies that PS-NcAF and TS-NcAF must have different symbol duration $T_s$, in order to keep the same information rate $R$.

With the above definitions, the average signal-to-noise ratios (SNRs) of the direct source-destination link, the source-relay link, and relay-destination link can be expressed, respectively, as\footnote{The parameters $\gamma_{0d}$ and $\gamma_{0r}$ are defined as the actual link SNRs associated with the direct link and the first-hop link, respectively; but $\gamma_{rd}$ is defined as the average SNR of the second-hop when the received signal from the first-hop is of unit energy, i.e., $\gamma_{rd}\pardef \frac{G_r^2\mathcal{L}_{rd}}{\sigma_{rd}^2}$.}
\begin{subequations}\label{eq:SNRs}
\begin{align}
   \gamma_{0d} & \pardef \left \{
   \begin{array}{ll}
    \frac{P_0\mathcal{L}_{0d}\log_2M}{(K+1)(\sigma_{0d,1}^2 + \sigma_{0d,2}^2)R}, & \text{PS-NcAF}, \\
    \frac{(1-\alpha)P_0\mathcal{L}_{0d}\log_2M}{(K+1)(\sigma_{0d,1}^2 + \sigma_{0d,2}^2)R}, & \text{TS-NcAF},
  \end{array}
   \right. \\
   \gamma_{0r} & \pardef \left \{
   \begin{array}{ll}
    \frac{(1-\rho)P_0\mathcal{L}_{0r}\log_2M}{(K+1)\big[(1-\rho)\sigma_{0r,1}^2+\sigma_{0r,2}^2\big]R}, & \text{PS-NcAF}, \\
    \frac{(1-\alpha)P_0\mathcal{L}_{0r}\log_2M}{(K+1)(\sigma_{0r,1}^2+\sigma_{0r,2}^2)R}, & \text{TS-NcAF},
  \end{array}
  \right. \\
   \gamma_{rd}& \pardef \left \{
   \begin{array}{ll}
    \frac{\rho \eta P_0\mathcal{L}_{0r}\mathcal{L}_{rd}\log_2M}{\big[(1-\rho)P_0\mathcal{L}_{0r}\log_2M+(K+1)\big((1-\rho)\sigma_{0r,1}^2+\sigma_{0r,2}^2\big)R\xi\big](\sigma_{rd,1}^2+\sigma_{rd,2}^2)}, & \text{PS-NcAF}, \\
    \frac{(K+1)\alpha \eta P_0\mathcal{L}_{0r}\mathcal{L}_{rd}\log_2M}{\big[(1-\alpha)P_0\mathcal{L}_{0r}\log_2M+(K+1)(\sigma_{0r,1}^2+\sigma_{0r,2}^2)R\xi\big](\sigma_{rd,1}^2+\sigma_{rd,2}^2)}, & \text{TS-NcAF},
  \end{array}
  \right.\label{eq:gamma_rd}
\end{align}
\end{subequations}
for $r = 1,\cdots,K$, where $\xi=1$ for $M$-DPSK and $\xi=M$ for $M$-FSK.
\subsection{Unified PS/TS-NcAF Framework for $M$-DPSK}
Following the parameter definitions in \eqref{eq:noise_variances}--\eqref{eq:SNRs}, the signal models for PS-NcAF and TS-NcAF employing $M$-DPSK can be expressed in a unified form as follows:
\begin{subequations}\label{eq:DPSK_unified}
\begin{align}
  \bm{y}_{0r}&=\sigma_{0r}\big(\sqrt{\gamma_{0r}}h_{0r}\bm{s}+\bm{n}_{0r}\big),\label{eq:y0r_DPSK_unified}\\
  \bm{y}_{0d}&=\sigma_{0d}\big(\sqrt{\gamma_{0d}}h_{0d}\bm{s}+\bm{n}_{0d}\big),\label{eq:y0d_DPSK_unified}\\
  \bm{y}_{rd}&=\sigma_{0r}\sigma_{rd}\sqrt{\gamma_{0r}\gamma_{rd}}h_{0r}h_{rd}\bm{s}_r+\sigma_{0r}\sigma_{rd}\sqrt{\gamma_{rd}}h_{rd}\bm{n}_{0r}+
  \sigma_{rd}\bm{n}_{rd},\label{eq:yrd_DPSK_unified}
\end{align}
\end{subequations}
where $\bm{n}_{ij}\sim\mathcal{CN}(\bm{0},\bm{I}_2)$ for $ij\in\{0d,0r,rd\}_{r=1}^K$. Note that the unified model in \eqref{eq:DPSK_unified} can represent either PS-NcAF or TS-NcAF by appropriately choosing the parameters according to \eqref{eq:noise_variances}--\eqref{eq:SNRs}.

\subsection{Unified PS/TS-NcAF Framework for Noncoherent $M$-FSK}
Following the definitions in \eqref{eq:noise_variances}--\eqref{eq:SNRs}, the signal models for PS-NcAF and TS-NcAF employing noncoherent $M$-FSK can be unified as follows:
\begin{subequations}\label{eq:FSK_unified}
\begin{align}
  \by_{0r}&=\sigma_{0r}\big(\sqrt{\gamma_{0r}}h_{0r}\bi_{m+1}+\bn_{0r}\big),\label{eq:y0r_FSK_unified}\\
  \by_{0d}&=\sigma_{0d}\big(\sqrt{\gamma_{0d}}h_{0d}\bi_{m+1}+\bn_{0d}\big),\label{eq:y0d_FSK_unified}\\
  \by_{rd}&=\sigma_{0r}\sigma_{rd}\sqrt{\gamma_{0r}\gamma_{rd}}h_{0r}h_{rd}\bi_{m+1}+\sigma_{0r}\sigma_{rd}\sqrt{\gamma_{rd}}h_{rd}\bn_{0r}+\sigma_{rd}\bn_{rd},\label{eq:yrd_FSK_unified}
\end{align}
\end{subequations}
where $\bn_{ij}\sim\mathcal{CN}(\bm{0},\bm{I}_M)$ for $ij\in\{0d,0r,rd\}$ and $r=1,\cdots,K$.
\begin{remark}
  The main benefit of the unified noncoherent SWIPT frameworks in \eqref{eq:DPSK_unified} and \eqref{eq:FSK_unified} lies in the fact that it enables unified design and analysis for the PS-NcAF and TS-NcAF protocols. For example, following these proposed unified noncoherent SWIPT frameworks, the noncoherent detectors for both PS-NcAF and TS-NcAF can be obtained in a unified form, which is treated in more detail in the next section.
\end{remark}

\section{Noncoherent Detectors for EH AF Networks}\label{sec:MLD}
In this section, the main objective is to develop the noncoherent detection schemes for EH AF relay systems. To proceed, we first tackle the mathematical challenges involved in the development of the noncoherent detectors. Then, the (exact) MLDs for PS-NcAF and TS-NcAF are derived in a unified form, which characterizes the optimum performance benchmark for noncoherent SWIPT in EH AF networks. Finally, low-complexity noncoherent detectors are derived, which achieves almost identical performance to the MLDs at dramatically lower complexity.

\subsection{Mathematical Preliminary}\label{sub:math}
The unified signal models in \eqref{eq:DPSK_unified} and \eqref{eq:FSK_unified} involve complicated transformations of the CSCG random variables/vectors. To facilitate the development of the ML detection schemes, it is useful to study the probability density functions (PDFs) of those Gaussian transformations.

\begin{lemma}\label{lemma:PDF}
Consider random variables $X_i\sim\mathcal{CN}(0,\Omega_i)$, $2\times 1$ random vectors $\bm{x}_i\sim\mathcal{CN}(\bm{0},\sigma_i^2\bm{I}_2)$, and $M\times 1$ random vectors $\bm{y}_i\sim\mathcal{CN}(\bm{0},\sigma_i^2\bm{I}_M)$, $i=1,2$, all of which are mutually independent. Let
\begin{align}
  \bm{X}_0&\pardef X_1X_2\bm{c}+X_2\bm{x}_1+\bm{x}_2,\label{eq:bm_X0}\\
  \bm{Y}_0&\pardef X_1X_2\bi_p+X_2\bm{y}_1+\bm{y}_2,\label{eq:bm_Y0}
\end{align}
where $p$ is any integer number between 1 and $M$, $\bm{c}\pardef [1,c]^T$, and $c$ is any complex number. The PDFs of $\bm{X}_0$, $f_{\bm{X}_0}(\bm{x})$, and $\bm{Y}_0$, $f_{\bm{Y}_0}(\bm{y})$, are given by
\begin{align}
  f_{\bm{X}_0}(\bm{x}) &= \frac{1}{(\pi\sigma_2^2)^2}I\bigg(\frac{\Omega_2\sigma_1^2}{\sigma_2^2},\Big[1+\frac{\Omega_1}{\sigma_1^2}\big(1+|c|^2\big)\Big]\frac{\Omega_2\sigma_1^2}{\sigma_2^2},
  \frac{|x_2-cx_1|^2}{(1+|c|^2)\sigma_2^2},\frac{|x_1+c^*x_2|^2}{(1+|c|^2)\sigma_2^2},1\bigg),\label{eq:PDF_X0}\\
  f_{\bm{Y}_0}(\bm{y}) &= \frac{1}{(\pi\sigma_2^2)^M}I\bigg(\frac{\Omega_2\sigma_1^2}{\sigma_2^2},\Big(1+\frac{\Omega_1}{\sigma_1^2}\Big)\frac{\Omega_2\sigma_1^2}{\sigma_2^2},
  \frac{\|\bm{y}\|^2-|y_p|^2}{\sigma_2^2},\frac{|y_p|^2}{\sigma_2^2},M-1\bigg),\label{eq:PDF_Y0}
\end{align}
where $\bm{x}\pardef [x_1,x_2]^T$, $\bm{y}\pardef [y_1,\cdots,y_M]^T$, and $I(\epsilon_1,\epsilon_2,\beta_1,\beta_2,\lambda)$ is defined as
\begin{align}
  I(\epsilon_1,\epsilon_2,\beta_1,\beta_2,\lambda)\pardef\int_0^\infty\frac{\e^{-\big(x+\frac{\beta_1}{1+\epsilon_1 x}+\frac{\beta_2}{1+\epsilon_2 x}\big)}}{(1+\epsilon_2x)^\lambda(1+\epsilon_2x)}\diff x,\label{eq:integral}
\end{align}
for any $\epsilon_1>0,\epsilon_2>0,\beta_1>0,\beta_2>0,\lambda>0$.
\end{lemma}
 \begin{proof}
   See Appendix \ref{app:PDF}.
\end{proof}

The PDF analysis in Lemma \ref{lemma:PDF} is very useful for finding the MLDs for EH AF relay systems, which will be addressed in the next subsection. Furthermore, the generic analytical results obtained in Lemma \ref{lemma:PDF} may be useful for other different applications involving the same Gaussian transformations as considered here.

\subsection{Unified Noncoherent MLDs}
Following the unified noncoherent SWIPT frameworks in \eqref{eq:DPSK_unified} and \eqref{eq:FSK_unified} and the generic PDF analysis in Lemma \ref{lemma:PDF}, in this subsection, the MLDs for PS-NcAF and TS-NcAF are obtained in a unified form.

\begin{theorem}\label{thm:Exact_MLD}
The MLDs for both PS-NcAF and TS-NcAF employing $M$-DPSK  can be expressed in a unified form as
\begin{align}
 \hat{m} = &\arg \max_{m=0,\cdots,M-1} \Bigg\{\frac{2\gamma_{0d}}{1+2\gamma_{0d}}\frac{\Re\{y_{0d}(l-1)y_{0d}^*(l)\e^{j2\pi m/M}\}}{\sigma_{0d}^2}
   +\sum_{r=1}^K\ln I\bigg(\sigma_{0r}^2\gamma_{rd}, (1+2\gamma_{0r})\sigma_{0r}^2\gamma_{rd},  \nonumber\\
   &~~~~~~~~~~~~~~~~~~~ \frac{|y_{rd}(l)-y_{rd}(l-1)\e^{j2\pi m/M}|^2}{2\sigma_{rd}^2},\frac{|y_{rd}(l)+y_{rd}(l-1)\e^{j2\pi m/M}|^2}{2\sigma_{rd}^2},1\bigg)\Bigg\}.\label{eq:MLD_DPSK}
\end{align}
Similarly, the MLDs for both PS-NcAF and TS-NcAF employing $M$-FSK are given in a unified form as
\begin{align}
  \hat{m} = &\arg\max_{m=0,\cdots,M-1}\Bigg\{\frac{\gamma_{0d}}{1+\gamma_{0d}}\frac{|\mathrm{y}_{0d}(m+1)|^2}{\sigma_{0d}^2}+\sum_{r=1}^K\ln I\bigg(\sigma_{0r}^2\gamma_{rd},(1+\gamma_{0r})\sigma_{0r}^2\gamma_{rd}, \nonumber\\
  &~~~~~~~~~~~~~~~~~~~~\frac{\|\by_{rd}\|^2-|\mathrm{y}_{rd}(m+1)|^2}{\sigma_{rd}^2},\frac{|\mathrm{y}_{rd}(m+1)|^2}{\sigma_{rd}^2},M-1\bigg)\Bigg\}.
   \label{eq:MLD_FSK}
\end{align}
\end{theorem}
\begin{proof}
See Appendix \ref{app:Exact_MLD}.
\end{proof}

The MLDs in \eqref{eq:MLD_DPSK} and \eqref{eq:MLD_FSK} involve the computations of the integral $I(\epsilon_1,\epsilon_2,\beta_1,\beta_2,\lambda)$ in \eqref{eq:integral}, which results in a high computational complexity. Nevertheless, the MLDs are very useful as they characterize the optimum performance benchmark for noncoherent SWIPT in EH AF networks. For example, one may use the MLDs as a theoretical performance upper bound  to test the performance of any suboptimum detectors, if developed in the literature.

\subsection{Unified Noncoherent GLDs}
In this subsection, to avoid integral evaluations and make the MLDs suitable for practical implementation, we adopt the Gauss-Legendre (GL) quadrature to approximate the integral, resulting in the so-called Gauss-Legendre detectors (GLDs) in closed-form.

\begin{theorem}\label{th:Approx_MLD}
  The GLDs for both PS-NcAF and TS-NcAF employing $M$-DPSK are given in the same form as \eqref{eq:MLD_DPSK}, except that the multivariate function $I(\cdot,\cdot,\cdot,\cdot,\cdot)$ is replaced by $\tilde{I}(\cdot,\cdot,\cdot,\cdot,\cdot)$, which is given by
   \begin{align}
    \tilde{I}(\epsilon_1,\epsilon_2,\beta_1,\beta_2,\lambda)\pardef \frac{1}{2}\sum_{i=1}^5w_i\psi\left(\frac{1+z_i}{2}\right).\label{eq:integral_app}
  \end{align}
  The parameters $w_i$ and $z_i$ are, respectively, the weights and nodes of the GL quadrature of order 5, where $w_1=128/225$, $w_2=w_3=\frac{322+13\sqrt{70}}{900}$, $w_4=w_5=\frac{322-13\sqrt{70}}{900}$, $z_1=0$, $z_2=-z_3=\frac{1}{3}\sqrt{5-2\sqrt{10/7}}$, $z_4=-z_5=\frac{1}{3}\sqrt{5+2\sqrt{10/7}}$, and $\psi(z)$ is defined as follows:
\begin{equation}\label{eq:Int_approx_subfunc}
  \psi(z)\pardef \frac{\e^{-\big(\frac{\beta_1}{1-\epsilon_1\ln z}+\frac{\beta_2}{1-\epsilon_2\ln z}\big)}}{(1-\epsilon_1\ln z)^\lambda(1-\epsilon_2\ln z)}.
\end{equation}
Similarly, The GLDs for PS-NcAF and TS-NcAF employing $M$-FSK are given in \eqref{eq:MLD_FSK}, where $I(\cdot,\cdot,\cdot,\cdot,\cdot)$ is replaced by $\tilde{I}(\cdot,\cdot,\cdot,\cdot,\cdot)$.
\end{theorem}
\begin{proof}
See Appendix \ref{app:Approx_MLD}.
\end{proof}

The MLDs in Theorem \ref{th:Approx_MLD} require the computation of $M$ integrals for detecting one information-bearing symbol, whereas the proposed GLDs avoid the integral computations and involve no special functions. Thus, the GLDs can be easily implemented in practice. Furthermore, we will demonstrate in the next section that the GLDs achieve almost the same error performance as the MLDs. Thus, the GLDs constitute practical solutions for EH noncoherent relaying systems.

\section{Numerical Results}\label{sec:numerical}
In this section, we evaluate the performance of the proposed noncoherent SWIPT schemes from various aspects through Monte Carlo simulations.

A $K$-relay line network model is considered, where the relays are distributed over the straight line between the source and destination. Given the source-destination distance $D_{0d}$,  the source-relay distance $D_{0r}$ and the relay-destination distance $D_{rd}$ may vary arbitrarily while guaranteeing $D_{0r}+D_{rd}=D_{0d}$,  for $r=1,\cdots,K$. The EH efficiency is set to $\eta = 0.6$ as in \cite{X.Zhou2013.11}. For ease of simulations, the antenna noise and ID circuit noise are assumed to have equal variances, i.e., $\sigma_{ij,1}^2=\sigma_{ij,2}^2\pardef\sigma_0^2/2$, $ij\in\{0d,0r,rd\}_{r=1}^K$, which ensures that each receiving node has a total noise variance of $\sigma_0^2$. Since the transmit power $P_0$ applied to the source is the only energy supply to the whole network, the error performance of the whole system is parameterized by $\mathrm{SNR}\pardef P_0/\sigma_0^2$. Unless otherwise stated, the communication distance is set to $D_{0d}=3$ (m) and the bounded path-loss model \cite{Z.Ding2014.8} is adopted, i.e., $\mathcal{L}_{ij}=\frac{1}{1+D_{ij}^\varrho}$, where the path loss exponent is $\varrho=4$ as in \cite{I.Krikidis2014.3}. We will evaluate the SER performance of the MLDs and GLDs for the proposed PS/TS-NcAF framework. In addition, the PS/TS-NcDF framework and the corresponding MLDs developed in \cite{P.Liu2014.12.submitted} will be considered  as a benchmark for the purpose of comparison.

\subsection{Comparison of PS-NcAF and TS-NcAF}

\begin{figure}[!t]%
\centering
\includegraphics[width=0.5\columnwidth]{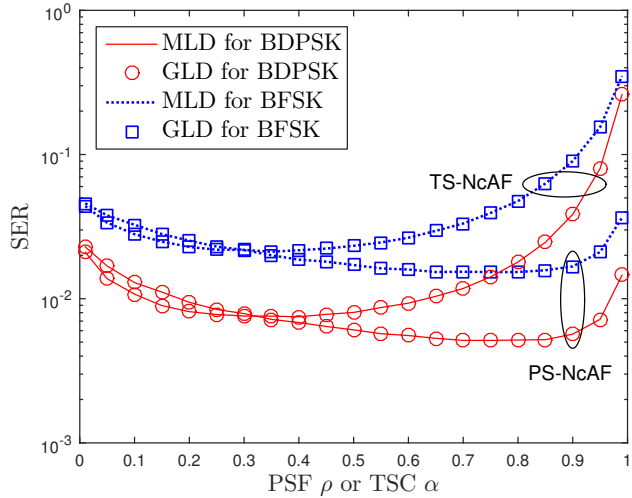}
\caption{SER versus PSF ($\rho$) or TSC ($\alpha$) for PS-NcAF and TS-NcAF employing BDPSK and BFSK at information rate $R=1$ bps and $\mathrm{SNR} = 35$ dB in a single-relay network ($K=1$) with $D_{0r}=2$.}
\label{fig:PS_TS_vary_coeff_35dB_M2_K1}
\end{figure}

%
%

\begin{figure*}[!t]%
\centering
\subfigure[][]{\includegraphics[width=0.5\columnwidth]{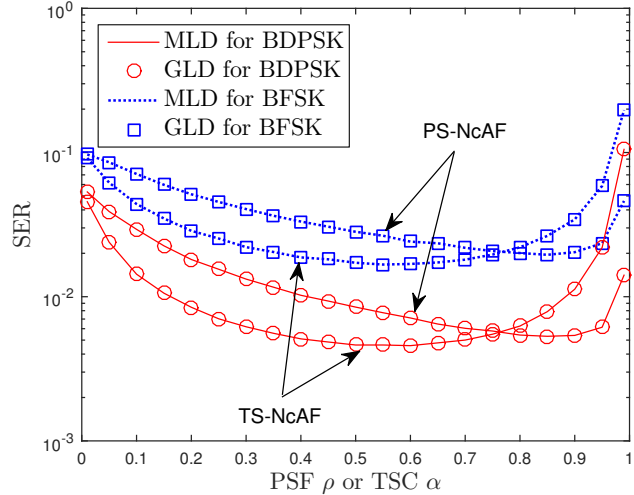}\label{fig:PS_TS_vary_coeff_30dB_M2_K2}} \hfil
\subfigure[][]{\includegraphics[width=0.5\columnwidth]{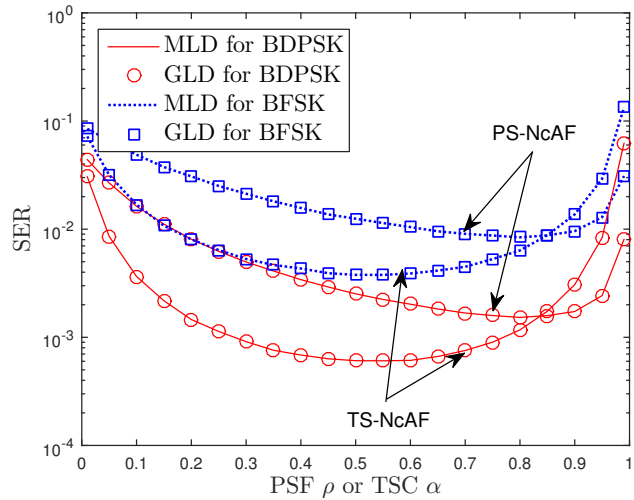}\label{fig:PS_TS_vary_coeff_35dB_M2_K3}}
\caption{SERs of PS/TS-NcAF and PS/TS-NcDF for multi-relay networks employing binary modulations. \subref{fig:PS_TS_vary_coeff_30dB_M2_K2} 2-relay case ($K=2$) with $\{D_{0r}\}_{r=1}^2=\{1,1.5\}$, $R=0.5$ bps, and $\mathrm{SNR} = 30$ dB. \subref{fig:PS_TS_vary_coeff_35dB_M2_K3} 3-relay case ($K=3$) with $\{D_{0r}\}_{r=1}^3=\{0.75,1.5,2.25\}$, $R=1$ bps,  and $\mathrm{SNR} = 35$ dB.}
\label{fig:PS_TS_higer_constellations}
\end{figure*}

In this subsection, we compare the error performance of PS-NcAF and TS-NcAF by investigating the impacts of the PSF $\rho$, the TSC $\alpha$, and the number of relays $K$. The comparisons will be carried out for binary DPSK (BDPSK) and binary FSK (BFSK).

\subsubsection{Impacts of the PSF and TSC}
The SER performance of the proposed EH relaying protocols is evaluated by varying the PSF $\rho$ for PS-NcAF and the TSC $\alpha$ for TS-NcAF while fixing the SNR as $\mathrm{SNR}=35$ dB for binary ($M=2$) noncoherent signalings at the information rate $R=1$ bps. Fig. \ref{fig:PS_TS_vary_coeff_35dB_M2_K1} illustrates the SERs versus the PSF $\rho$ or TSC $\alpha$ for the \emph{single-relay} ($K=1$) case where the relay is located at $D_{0r} =1$. We see that there exist (unique) optimum values of the PSF (i.e., $\rho=0.8$) and the TSC (i.e., $\alpha=0.4$) which minimize the SERs of PS-NcAF and TS-NcAF, respectively. This is because the choices of the PSF or TSC result in some tradeoff between EH and ID in EH relay systems. Specifically, for larger $\rho$ (or $\alpha$), more energy can be harvested and utilized for data relaying, whereas less signal energy (or time) can be used for ID. For smaller $\rho$ (or $\alpha$), more signal energy (or time) is available for ID; but less energy can be harvested and used for data relaying. Due to this tradeoff, the optimum values of $\rho$ and $\alpha$ always lie between 0 and 1, which balances the operation of EH and ID such that the overall SER is minimized.


\begin{figure}[!t]%
\centering
\includegraphics[width=0.6\columnwidth]{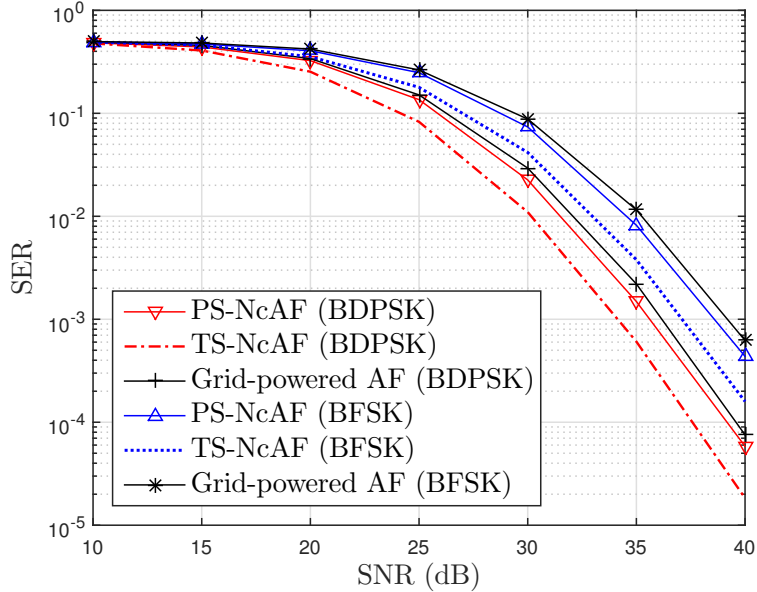}
\caption{Grid-powered relay system with power $P_0/(K+1)$ allocated to each terminal versus EH relay system with $P_0$ supplied to the source only for binary noncoherent signalings with $R=1$ bps, where $K=3$, $\{D_{0r}\}_{r=1}^3=\{0.75,1.5,2.25\}$, $\rho=0.8$, and $\alpha=0.55$.}
\label{fig:Grid_power_AF_K3_total_P0}
\end{figure}

\subsubsection{Impact of the number of relays}
For the single-relay network as illustrated in Fig. \ref{fig:PS_TS_vary_coeff_35dB_M2_K1},  the minimum SER of PS-NcAF (achieved around $\rho= 0.8$) is lower than the minimum SER of TS-NcAF (achieved around $\alpha=0.4$). However, this relationship is reversed in the \emph{multi}-relay case. Specifically, for the 2-relay case ($K=2$) employing binary noncoherent signalings where $\{D_{0r}\}_{r=1}^3=\{1,1.5\}$ and $R=0.5$ bps, Fig. \ref{fig:PS_TS_vary_coeff_30dB_M2_K2} shows
that the minimum SER of TS-NcAF achieved around $\alpha=0.6$ is slightly smaller than the minimum SER of PS-NcAF achieved around $\rho=0.85$, for both BDPSK and BFSK. In addition, for the 3-relay  ($K=3$) case where $\{D_{0r}\}_{r=1}^3=\{0.75,1.5,2.25\}$ and $R=1$ bps, as shown in Fig. \ref{fig:PS_TS_vary_coeff_35dB_M2_K3}, we can clearly see that the minimum SER of TS-NcAF achieved around $\alpha=0.55$ is much smaller than the minimum SER of PS-NcAF achieved around $\rho=0.8$. These comparisons indicate that the number of relays $K$ is a key factor which dictates to the superior EH protocol. Specifically, PS-NcAF outperforms TS-NcAF in the single-relay case ($K=1$), while TS-NcAF outperforms PS-NcAF in the multi-relay case ($K\geq 2$). The reason for this is explained as follows. For PS-NcAF, since the EH time and the ID time are equal (both equal to $\frac{T}{K+1}$), the harvested power in \eqref{eq:Pk_PS} for data relaying at each relay is independent of $K$, which makes the second-hop SNR $\gamma_{rd}$ independent of $K$ as well, as illustrated in \eqref{eq:gamma_rd}. In contrast, for TS-NcAF, due to the time switching operation, the EH time, $\alpha T$, is generally unequal to the ID time, $\frac{(1-\alpha)T}{K+1}$, at each relay. Thus, for fixed harvested energy, the harvested power for data relaying at each relay becomes proportional to $K+1$, and consequently, the second-hop SNR $\gamma_{rd}$ is a monotonically increasing function of $K$, as illustrated in \eqref{eq:gamma_rd}. Hence, as $K$ increases, the second-hop link for TS-NcAF becomes increasingly more reliable as compared to that for PS-NcAF, which results in extra performance improvement over PS-NcAF for $K\geq 2$.


For the multi-relay case, we are also interesting in comparing TS-NcAF and PS-NcAF at different SNRs, such that the performance gain of TS-NcAF over PS-NcAF can be quantified at any target SERs. Fig. \ref{fig:Grid_power_AF_K3_total_P0} shows the SERs versus the SNR for the 3-relay network employing the optimum EH parameters, i.e., $\rho = 0.8$ for PS-NcAF and $\alpha=0.55$ for TS-NcAF according to Fig. \ref{fig:PS_TS_vary_coeff_35dB_M2_K3}, where $\{D_{0r}\}_{r=1}^3=\{0.75,1.5,2.25\}$ and $R=1$ bps. For the target SER of $10^{-3}$, we see that TS-NcAF outperforms PS-NcAF by about 1.5 dB, for both BDPSK and BFSK. This significant gain is very attractive as the energy efficiency is of major concern in EH systems. Furthermore, the conventional grid-powered relay system is evaluated as a benchmark. For a fair comparison in terms of the total power consumption, the  powers allocated to the source and all relays are equal to $\frac{P_0}{K+1}$, ensuring the same total power of $P_0$ as for the EH relay system. We observe that both EH protocols, namely PS-NcAF and TS-NcAF, outperform the grid-powered AF relaying protocol, for the same noncoherent modulation. This is because given the same total power, the EH ability allows the relays to obtain extra powers from the ambient RF signals, thus boosting the overall performance over the grid-powered relay system with no EH.

\subsection{Impacts of Relay Positions and the Path Loss Exponent}

\begin{figure*}[!t]%
\centering
\subfigure[][]{\includegraphics[width=0.6\columnwidth]{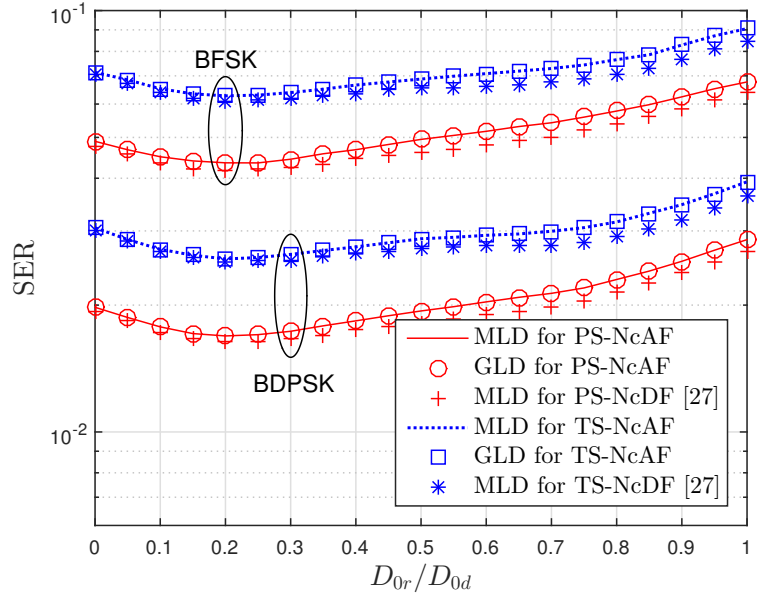}\label{fig:PS_TS_vary_d_26dB_2p7}} \hfil
\subfigure[][]{\includegraphics[width=0.6\columnwidth]{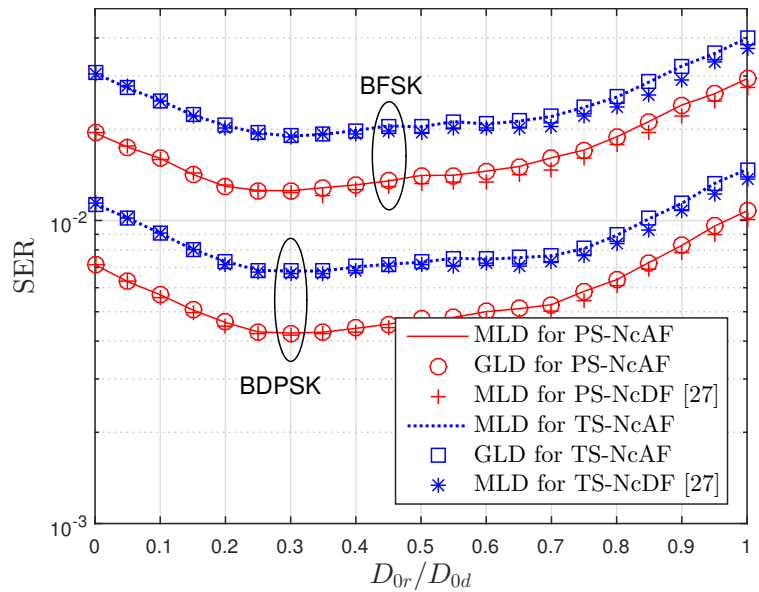}\label{fig:PS_TS_vary_d_35dB_4}}
\caption{SER versus the relative relay position ($D_{0r}/D_{0d}$) for PS/TS-NcAF and PS/TS-NcDF in the single-relay case with PSF $\rho = 0.8$, TSC $\alpha=0.4$, and information rate $R=1$ bps. \subref{fig:PS_TS_vary_d_26dB_2p7} $\varrho=2.7$ and $\mathrm{SNR}=26$ dB. \subref{fig:PS_TS_vary_d_35dB_4} $\varrho=4$ and $\mathrm{SNR} = 35$ dB.}
\label{fig:PS_TS_vary_d}
\end{figure*}

\subsubsection{Impact of the relay position $D_{0r}$}
For conventional self-powered relay systems where the EH capability is disabled \cite{A.Ribeiro2005.5,M.R.Souryal2006.5,Y.Zhu2010.1,D.Chen2006.7,M.R.Souryal2010.7, W.Cho2008.11}, the impact of the relay position on the error performance is well understood.  Specifically, the optimum relay position minimizing the SER in \emph{coherent} relay systems is at the midpoint between the source and destination, for both AF and DF \cite{A.Ribeiro2005.5,M.R.Souryal2006.5}. For \emph{noncoherent} relay systems, however, the optimum relay positions differ  for AF and DF. For DF relaying, the optimum relay position is closer to the source than to the destination \cite{Y.Zhu2010.1,D.Chen2006.7}, whereas the optimum relay position for AF relaying is closer to the destination than to the source \cite{M.R.Souryal2010.7, W.Cho2008.11}.

Fig. \ref{fig:PS_TS_vary_d} illustrates the SER versus the relative relay position $D_{0r}/D_{0d}$ for noncoherent EH relay systems. For both AF-based PS/TS-NcAF proposed in this paper and the DF-based PS/TS-NcDF in \cite{P.Liu2014.12.submitted}, the optimum relay positions are $D_{0r}=0.2D_{0d}$ when the path loss exponent is $\varrho=2.7$, as shown in Fig. \ref{fig:PS_TS_vary_d_26dB_2p7}, and $D_{0r}=0.3D_{0d}$ when the path loss exponent is $\varrho=4$, as shown in Fig. \ref{fig:PS_TS_vary_d_35dB_4}. That is, unlike the conventional noncoherent relay networks where the optimum relay positions depend on the relaying protocols (AF or DF), the optimum relay positions for noncoherent EH relay systems are invariant with respect to AF and DF, and are always closer to the source than to the destination. This major difference is due to the fact that EH relays are solely powered by the source, and thus, the dominant performance limiting factor in EH relay networks is the harvested energy at the relays, which is invariant with respect to AF and DF. Moreover, the overall performance is affected by the path loss, which is a distance-dependent constant. Since both the EH operation and the path loss effect are irrelevant to the relaying protocols, the optimum relay positions are the same with respect to AF and DF in EH relay systems. We note that the EH operation and the path loss have different impacts on the relay position. On one hand, the relays must be sufficiently close to the source in order to harvest enough energy. On the other hand, the relays must not be infinitely close to the source (e.g., located at the source) because otherwise the relay-to-destination path loss would become very significant, which degrades the overall performance. Considering both the EH and path-loss effects, the optimum relay position for EH relay systems is relatively closer to the source than to the destination.

\subsubsection{Impact of the path loss exponent $\varrho$}
As illustrated in Fig. \ref{fig:PS_TS_vary_d}, the optimum relay position, which is always closer to the source than to the destination, shifts slightly towards the destination as the path loss exponent $\varrho$ increases from 2.7 to 4. This is because a larger path loss exponent corresponds to a larger relay-to-destination path loss, which can degrade the overall error performance. As such, the relay should step towards the destination a little bit to diminish the effect of the increased path loss. Nevertheless, the prerequisite of EH operation cannot be obviated, and thus, the relays must always be deployed closer to the source than to the destination, with some slight movements towards the destination as the path loss exponent increases.

\subsection{Impact of Higher Constellations}

%
%

\begin{figure*}[!t]%
\centering
\subfigure[][]{\includegraphics[width=0.6\columnwidth]{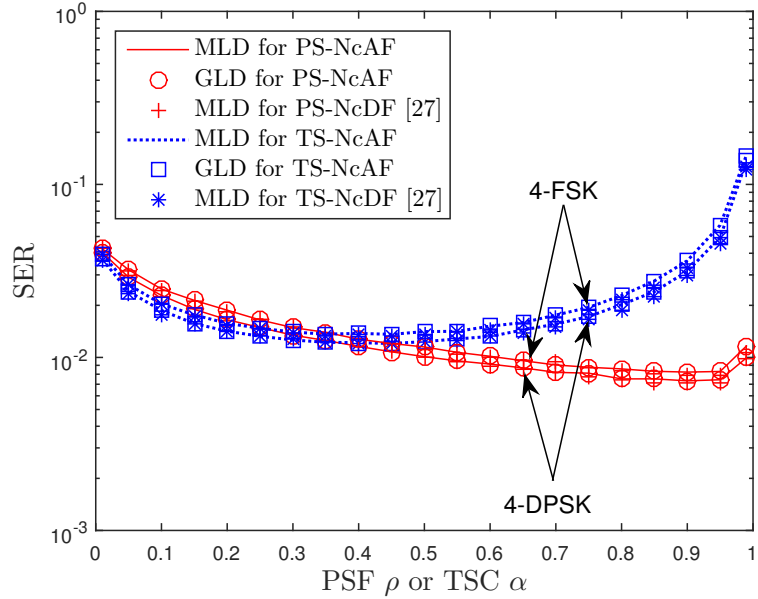}\label{fig:PS_TS_vary_coeff_38dB_M4_K1}} \hfil
\subfigure[][]{\includegraphics[width=0.6\columnwidth]{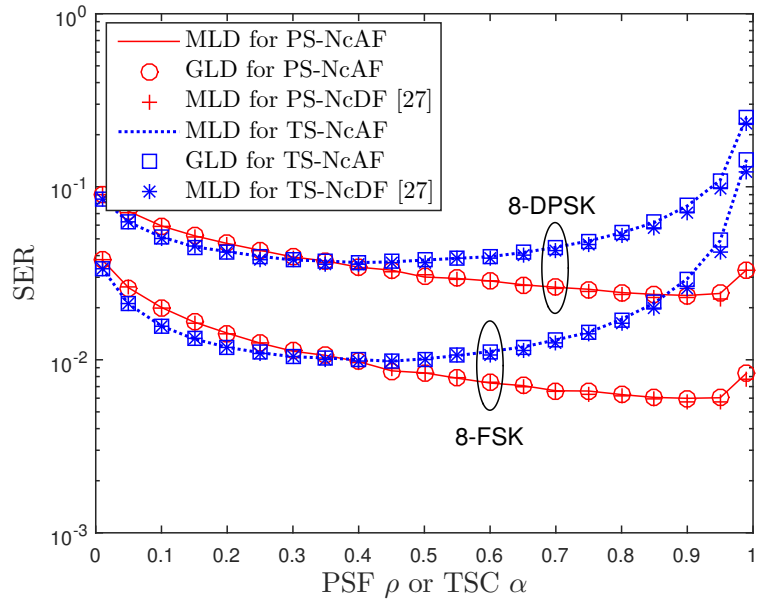}\label{fig:PS_TS_vary_coeff_40dB_M8_K1}}
\caption{SERs of PS/TS-NcAF and PS/TS-NcDF for higher-order constellations in the single-relay case ($K=1$) with $D_{0r}=1$. \subref{fig:PS_TS_vary_coeff_38dB_M4_K1} $M=4$, $\mathrm{SNR} = 38$ dB, and $R=2$ bps. \subref{fig:PS_TS_vary_coeff_40dB_M8_K1} $M=8$, $\mathrm{SNR} = 40$ dB, and $R=3$ bps.}
\label{fig:PS_TS_higer_constellations}
\end{figure*}

So far, we have only considered noncoherent binary  signalings with $M=2$, and a 3 dB performance gain of BDPSK over BFSK can be observed in Fig. \ref{fig:Grid_power_AF_K3_total_P0}. In this subsection, we will evaluate the performance of higher order noncoherent signalings for $M>2$. Specifically, for a single-relay network with $D_{0r}=1$, the SER of $M=4$ at $\mathrm{SNR} = 38$ dB and $R=2$ bps is illustrated in Fig. \ref{fig:PS_TS_vary_coeff_38dB_M4_K1}, where 4-DPSK slightly outperforms 4-FSK, irrespective of the EH relaying protocols (e.g., PS/TS-NcAF or PS/TS-NcDF). Moreover, the SER for $M=8$ at $\mathrm{SNR} = 40$ dB and $R=3$ bps is illustrated in Fig. \ref{fig:PS_TS_vary_coeff_40dB_M8_K1}, which shows that 8-FSK substantially outperforms 8-DPSK, irrespective of the EH relaying protocols. Observing the trend from $M=2$ to 4 and 8, we conclude that $M$-FSK is more efficient in terms of energy consumption than $M$-DPSK, which results in significant performance gain for $M\geq 8$. In particular, the performance gain of 8-FSK over 8-DPSK is about 4 dB in the target SER of $10^{-2}$, as demonstrated in \cite{P.Liu2014.12.submitted}.

\subsection{Comparison of PS/TS-NcAF with PS/TS-NcDF \cite{P.Liu2014.12.submitted}}

\begin{figure}[!t]%
\centering
\includegraphics[width=0.6\columnwidth]{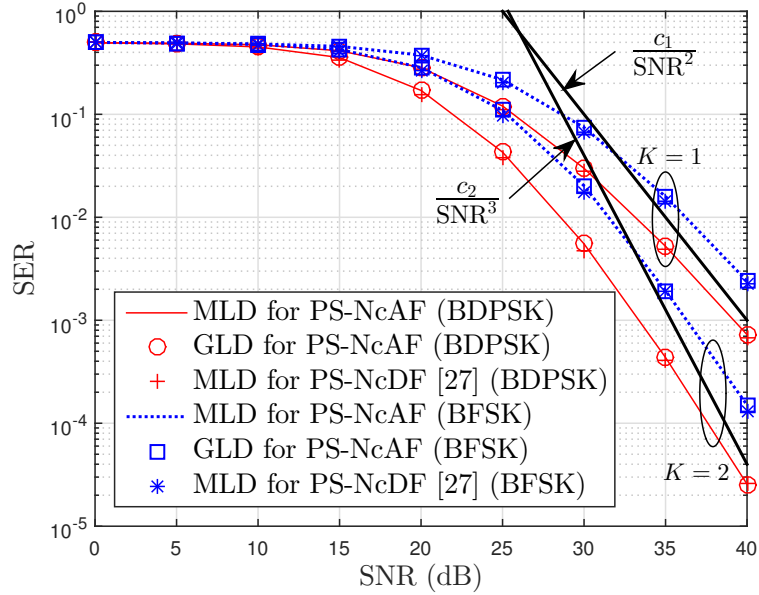}
\caption{SER versus SNR for PS-NcAF and PS-NcDF employing BDPSK and BFSK: 1) single-relay ($K=1$) case with $D_{0r}=2$,  $\rho=0.8$, and $R=1$ bps and 2) two-relay ($K=2$) case with $\{D_{0r}\}_{r=1}^2=\{1,1.5\}$,  $\rho=0.85$, and $R=0.5$ bps.}
\label{fig:PS_NcAF_validation}
\end{figure}

\begin{figure}[!t]%
\centering
\includegraphics[width=0.6\columnwidth]{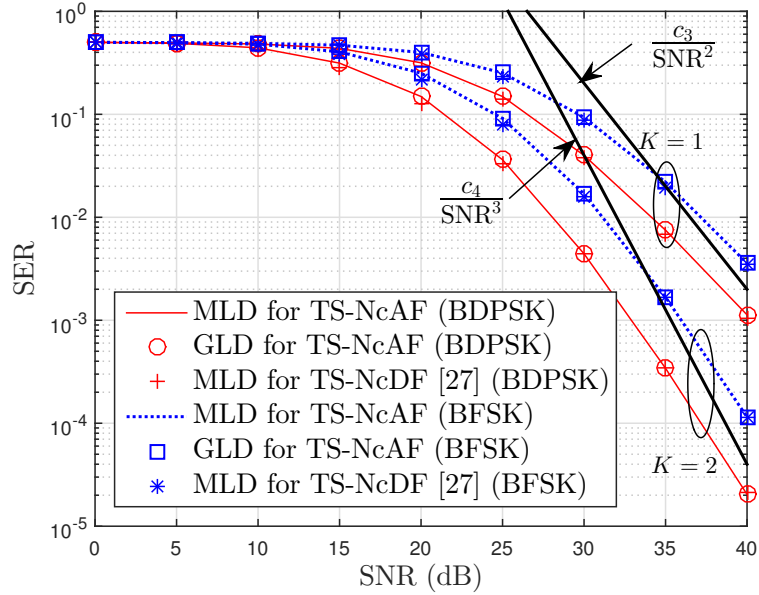}
\caption{SER versus SNR for TS-NcAF and TS-NcDF employing BDPSK and BFSK: 1) single-relay ($K=1$) case with $D_{0r}=2$,  $\alpha=0.4$, and $R=1$ bps and 2) two-relay ($K=2$) case with $\{D_{0r}\}_{r=1}^2=\{1,1.5\}$,  $\alpha=0.6$, and $R=0.5$ bps.}
\label{fig:TS_NcAF_validation}
\end{figure}

Since both AF and DF relaying protocols are useful candidates for achieving SWIPT in EH relay systems, it is meaningful to compare the performance of the AF-based PS/TS-NcAF proposed in this paper with that of the DF-based PS/TS-NcDF developed in \cite{P.Liu2014.12.submitted}. The SER performance of the EH relaying protocols generally depend on the relay positions, the EH parameters (such as PSF and TSC), and SNRs. Thus, performance comparisons can be carried out from various angles. In particular, Fig. \ref{fig:PS_TS_vary_d} illustrates the comparison of PS/TS-NcAF and PS/TS-NcDF by varying the relay positions for fixed EH parameters and fixed SNR. In addition, Fig. \ref{fig:PS_TS_higer_constellations} illustrates the SERs for a variety of EH parameter settings (PSF or TSC), where the relay position and SNR are fixed. The remaining possible comparison is therefore to fix the relay position and EH parameters, while varying the SNRs. To this end, we compare PS-NcAF with PS-NcDF, and TS-NcAF with TS-NcDF at different SNRs in Figs. \ref{fig:PS_NcAF_validation}  and  \ref{fig:TS_NcAF_validation}, respectively, where two typical network geometries are considered: 1) single-relay ($K=1$) case with $D_{0r}=2$,  $\alpha=0.4$, $\rho=0.8$, and $R=1$ bps and 2) two-relay ($K=2$) case with $\{D_{0r}\}_{r=1}^2=\{1,1.5\}$,  $\alpha=0.6$, $\rho=0.85$, and $R=0.5$ bps. Note that the PSF $\rho$ and TSC $\alpha$ employed in the comparisons of Figs. \ref{fig:PS_NcAF_validation} and \ref{fig:TS_NcAF_validation} are optimum values determined in Figs. \ref{fig:PS_TS_vary_coeff_35dB_M2_K1} and \ref{fig:PS_TS_vary_coeff_30dB_M2_K2}. For all the comparisons carried out from various angles in Figs. \ref{fig:PS_TS_vary_d}--\ref{fig:TS_NcAF_validation}, we observe that the SER curves associated with PS-NcAF and PS-NcDF almost overlap, so do the SER curves associated with TS-NcAF and TS-NcDF. These observations reveal that the noncoherent AF and DF relaying protocols yield almost the same error performance in EH relay networks, regardless of the SWIPT architectures (e.g., power splitting and time switching), the modulation types (e.g., DPSK and FSK), and system parameters (e.g., information rate $R$, number of relays $K$,  relay positions, and SNRs). This is in contrast to the conventional self-powered relay networks, where either AF or DF may outperform each other depending on the relay positions. In particular, for self-powered relay networks, DF might outperform AF when the relay is much closer to the source than to the destination, because the noise amplification due to AF substantially degrades the performance. On the other hand, AF might outperform DF when the relay is much closer to the destination, due to the high relay decoding error probabilities in DF. However, both effects of noise amplification (associated with AF) and error propagation (associated with DF) are significantly weakened and become less important in EH relay systems, because the dominant performance limiting factor is the harvested energy at the relay, which is invariant with respect to the AF and DF relaying protocols. Hence, for EH relay systems, there is almost no difference between the noncoherent AF and DF relaying in terms of the error performance.

In Figs. \ref{fig:PS_NcAF_validation} and \ref{fig:TS_NcAF_validation}, we also observe that the proposed noncoherent detectors achieve full diversity orders of 2 and 3 for the single- and two-relay networks, respectively, with the direct link. Furthermore, we would like to note that the SER performance of the GLD is in excellent agreement with the SER performance of the MLD, as demonstrated in all Figs. \ref{fig:PS_TS_vary_coeff_35dB_M2_K1}--\ref{fig:TS_NcAF_validation}. Moreover, the GLD is given in closed-form, involving no integrals or special functions at all. This suggests that the closed-form GLD is an appealing practical solution for noncoherent SWIPT in EH AF relay systems. Finally, this also indicates that $\tilde{I}(\epsilon_1,\epsilon_2,\beta_1,\beta_2,\lambda)$ of \eqref{eq:integral_app} is a very accurate approximation for the integral   $I(\epsilon_1,\epsilon_2,\beta_1,\beta_2,\lambda)$ in \eqref{eq:integral}.
 
\subsection{Performance Evaluation for Practical Indoor Communications Scenario}
\begin{figure}[!t]%
\centering
\includegraphics[width=0.6\columnwidth]{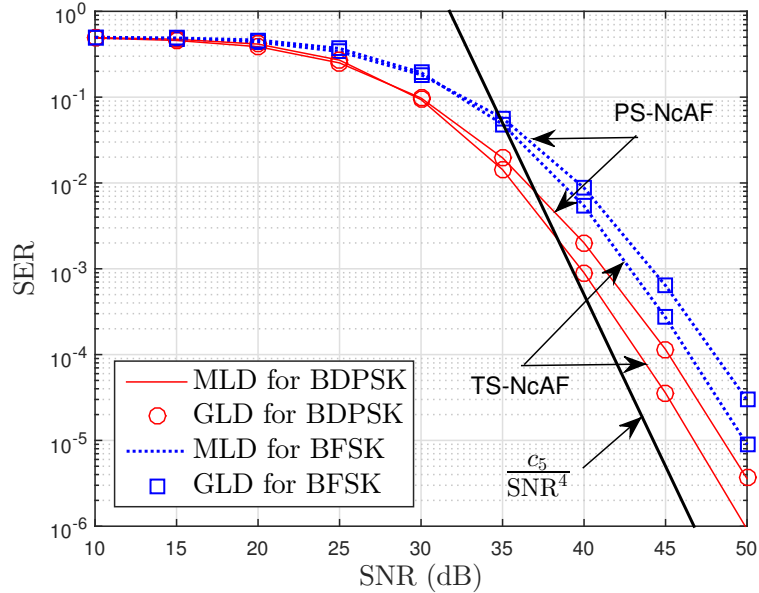}
\caption{SER performance for typical indoor communications with path loss $\mathcal{L}_{ij}~ \text{(dB)}=10\log_{10}\frac{1}{1+D_{ij}^{1.6}}-3.4$, where $D_{0d}=10$ (m),  $D_{0r}=0.1D_{0r}$ for $r=1,2,3$, $\alpha=0.55$, $\rho=0.8$, and $R=1$ bps.}
\label{fig:Indoor_3_relays}
\end{figure}

We now evaluate the performance of the proposed detectors under a practical indoor communication scenario operating at 900 MHz, where the typical  communication distance of $D_{0d}=10$ (m) is considered. Note that according to the current state-of-the-art of RF energy harvesting/transfer, the maximum line-of-sight (LOS) operating distance for some RF energy harvesting chips is 12--14 meters \cite{website:powercast,S.Bi2014.to_appear}. In this paper, we consider Rayleigh fading with blocked LOS due to a double plasterboard wall; thus, the communication distance of 10 meters is a very reasonable and practical choice. According to empirical measurements carried out at 900 MHz,, we set the path loss exponent as $\varrho=1.6$, which corresponds to the communication inside the typical office building  \cite[Table 2.2]{A.Goldsmith2005book}. The partition loss is set to 3.4 dB assuming that the transmit and receive terminals are separated by the double plasterboard wall \cite[Table 2.1]{A.Goldsmith2005book}. Also, we assume that the communication occurs on the same floor; thus, the floor loss can be ignored. According to the indoor path loss model featuring at 900 MHz \cite[eq. (2.38)]{A.Goldsmith2005book}, we set the path loss as $\mathcal{L}_{ij}~ \text{(dB)}=10\log_{10}\frac{1}{1+D_{ij}^{1.6}}-3.4$ in this subsection. Fig. \ref{fig:Indoor_3_relays} shows the SER performance for an EH relay system with $K=3$ relays, all located at $D_{0r}=0.1D_{0d}$, $r=1,2,3$. Note that we choose this relay position because the optimum relay position is much closer to the source than to the destination, as indicated in Fig. \ref{fig:PS_TS_vary_d}. Also, the EH parameters are chosen as $\alpha=0.55$ and $\rho=0.8$, which are optimum values for $K=3$ relays according to Fig. \ref{fig:PS_TS_vary_coeff_35dB_M2_K3}. Our simulation in Fig. \ref{fig:Indoor_3_relays} demonstrates that for binary modulations with $R=1$ bps, the diversity order of 4 is achieved for $K=3$ EH relays (with the direct link), which confirms the full diversity performance.

\section{Conclusions}\label{sec:con}
In this paper, we have developed a noncoherent SWIPT framework for EH AF relay systems, which embraces both PS-NcAF and TS-NcAF in a unified form and supports arbitrary $M$-ary noncoherent signalings including $M$-FSK and $M$-DPSK. The main advantage of the proposed SWIPT scheme is that it eliminates the need for the instantaneous CSI, which alleviates the system overhead and reduces the energy consumption.  Following this framework, we developed noncoherent MLDs for PS-NcAF and TS-NcAF in a \emph{unified} form, which involved integral evaluations yet served as the optimum performance benchmark for noncoherent SWIPT. To avoid integral computations, we also proposed \emph{closed-form} GLDs, which achieved almost identical SER performance to the MLDs at substantially lower computational complexity. It was demonstrated that these noncoherent detectors achieve full spatial diversity in Rayleigh fading. Moreover, in terms of the error performance, the proposed SWIPT relaying schemes may outperform conventional grid-powered relay systems under the same total power constraint. Finally, numerical results led to useful design insights into the noncoherent SWIPT in various aspects, including the effects of the time switching or power splitting parameters, relaying protocols, the number of relays, relay positions, and the modulation alphabet size, which are summarized as follows:
\begin{itemize}
  \item The choice of the power slitting factor or time switching coefficient represents some tradeoff between energy harvesting and information delivery. Unique optimal values of the PSF or TSC minimizing the SER exist between 0 and 1.
  \item The number of EH relays is a key factor on the performance of noncoherent SWIPT: PS-NcAF outperforms TS-NcAF in the single-relay case, whereas TS-NcAF outperforms PS-NcAF in the multi-relay case.
  \item The optimum relay position is closer to the source than to the destination, with some slight movements towards the destination as the path loss exponent increases.
  \item The noncoherent $M$-FSK with $M\geq 8$ is more energy-efficient  and thus might be more suitable than $M$-DPSK for EH relay systems, regardless of the EH relaying protocols.
  \item The noncoherent AF and DF relaying protocols yield almost the same error performance in EH relay networks, regardless of the SWIPT architectures, the modulation types, and system parameters.
\end{itemize}

An interesting extension of this work is to analyze the SER of the MLD or GLD, and using the obtained SER expression to \emph{analytically} determine the optimum time switching coefficient  or power splitting factor. In addition, more sophisticated and intelligent coordination of the source and relays' transmissions (such as relay selection) may also be considered in order to further boost the performance.

\appendices

\renewcommand{\thesection}{\Alph{section}}
\renewcommand{\theequation}{\Alph{section}.\arabic{equation}}
\setcounter{equation}{0}
\section{Proof of Lemma \ref{lemma:PDF}}\label{app:PDF}
In this proof, we will first study the distribution of $\bm{X}_0$, followed by a brief derivation of the PDF of $\bm{Y}_0$.

It is not hard to show that the conditional distribution of $\bm{X}_0$ given $|X_2|^2=z$ is $\bm{X}_0|_{|X_2|^2=z}\sim\mathcal{CN}(\bm{0},\bm{\Sigma}_0)$, where
\begin{equation}
  \bm{\Sigma}_0
  =\left[\begin{array}{cc}
     (\Omega_1+\sigma_1^2)z+\sigma_2^2, & \Omega_1c^*z \\
     \Omega_1cz, & (\Omega_1|c|^2+\sigma_1^2)z+\sigma_2^2
   \end{array}
    \right].
\end{equation}
We denote by $\det(\bm{\Sigma}_0)$ the determinant of $\bm{\Sigma}_0$, and it can be shown that
\begin{align}
  \det(\bm{\Sigma}_0)&=(\sigma_1^2z+\sigma_2^2)\big([\Omega_1(1+|c|^2)+\sigma_1^2]z+\sigma_2^2\big),\label{eq:det}\\
  \bm{\Sigma}_0^{-1}&=\frac{1}{\det(\mathrm{\Sigma_0})}\left[\begin{array}{cc}
     (\Omega_1|c|^2+\sigma_1^2)z+\sigma_2^2, & -\Omega_1c^*z \\
     -\Omega_1cz, & (\Omega_1+\sigma_1^2)z+\sigma_2^2
   \end{array}
    \right].\label{eq:inverse_matrix}
\end{align}
It follows that the conditional PDF of $\bm{X}_0$ given $|X_2|^2=z$ is $f_{\bm{X}_0\big|{|X_2|^2=z}}(\bm{x}) = \frac{1}{\pi^2\det(\bm{\Sigma}_0)}\e^{-\bm{x}^H\bm{\Sigma}_0^{-1}\bm{x}}$. Substituting in the expressions of \eqref{eq:det} and \eqref{eq:inverse_matrix}, we can show that
\begin{align}
  \bm{x}^H\bm{\Sigma}_0^{-1}\bm{x}&=\frac{1}{\det(\bm{\Sigma}_0)}[x_1^*,x_2^*]\left[\begin{array}{cc}
     (\Omega_1|c|^2+\sigma_1^2)z+\sigma_2^2, & -\Omega_1c^*z \\
     -\Omega_1cz, & (\Omega_1+\sigma_1^2)z+\sigma_2^2
   \end{array}
    \right]\left[\begin{array}{c}
     x_1 \\
     x_2
   \end{array}
    \right]\label{eq:app_matrix_prod1}\\
    &=\frac{[(\Omega_1|c|^2+\sigma_1^2)z+\sigma_2^2]|x_1|^2+[(\Omega_1+\sigma_1^2)z+\sigma_2^2]|x_2|^2-2\Omega_1z\Re(cx_1x_2^*)}{(\sigma_1^2z+\sigma_2^2)
    \big([\Omega_1(1+|c|^2)+\sigma_1^2]z+\sigma_2^2\big)}\label{eq:app_matrix_prod2}\\
    &=\frac{A_1}{\sigma_1^2z+\sigma_2^2}+\frac{A_2}{[\Omega_1(1+|c|^2)+\sigma_1^2]z+\sigma_2^2},\label{eq:app_matrix_prod3}
\end{align}
where \eqref{eq:app_matrix_prod3} follows by the partial fraction expansion. It can be shown that
\begin{align}
  A_1&=(\sigma_1^2z+\sigma_2^2)\bm{x}^H\bm{\Sigma}_0^{-1}\bm{x}\Big|_{z=-\frac{\sigma_2^2}{\sigma_1^2}}=\frac{|x_2-cx_1|^2}{1+|c|^2},\label{eq:app_A1}\\
  A_2&=\big([\Omega_1(1+|c|^2)+\sigma_1^2]z+\sigma_2^2\big)\bm{x}^H\bm{\Sigma}_0^{-1}\bm{x}\Big|_{z=-\frac{\sigma_2^2}{\Omega_1(1+|c|^2)+\sigma_1^2}}
     =\frac{|x_1+c^*x_2|^2}{1+|c|^2}.\label{eq:app_A2}
\end{align}
The unconditional PDF of $\bm{X_0}$ can be computed as follows:
\begin{align}
  f_{\bm{X}_0}(\bm{x})
 & =\int_0^\infty f_{\bm{X}_0\big|{|X_2|^2=z}}(\bm{x})f_{|X_2|^2}(z)\diff z  \label{eq:app_uncon_PDF_DPSK1}\\
 & = \frac{1}{\pi^2\Omega_2}\int_0^\infty \frac{\e^{-\big(\frac{z}{\Omega_2}+\frac{A_1}{\sigma_1^2z+\sigma_2^2}+\frac{A_2}{[\Omega_1(1+|c|^2)+\sigma_1^2]z+\sigma_2^2}\big)}}{(\sigma_1^2z+\sigma_2^2)
    \big([\Omega_1(1+|c|^2)+\sigma_1^2]z+\sigma_2^2\big)} \diff z.\label{eq:app_uncon_PDF_DPSK2}
 \end{align}
Applying the change of variable $\frac{z}{\Omega_2}\rightarrow t$ in \eqref{eq:app_uncon_PDF_DPSK2} and taking some algebraic manipulations, we can show that
\begin{align}
  f_{\bm{X}_0}(\bm{x})
 & = \frac{1}{(\pi\sigma_2^2)^2}\int_0^\infty \frac{\exp\bigg[-\Big(t+\frac{B_1}{1+\frac{\Omega_2\sigma_1^2}{\sigma_2^2}t}+\frac{B_2}{1+\frac{\Omega_2[\Omega_1(1+|c|^2)+\sigma_1^2]}{\sigma_2^2}t}\Big)\bigg]}
 {\left(1+\frac{\Omega_2\sigma_1^2}{\sigma_2^2}t\right)\left(1+\frac{\Omega_2[\Omega_1(1+|c|^2)+\sigma_1^2]}{\sigma_2^2}t\right)} \diff z,\label{eq:app_uncon_PDF_DPSK3}
 \end{align}
where
\begin{align}
  B_1\pardef \frac{A_1}{\sigma_2^2}=\frac{|x_2-cx_1|^2}{(1+|c|^2)\sigma_2^2},\label{eq:app_B1}\\
  B_2\pardef \frac{A_2}{\sigma_2^2}=\frac{|x_1+c^*x_2|^2}{(1+|c|^2)\sigma_2^2}.\label{eq:app_B2}
\end{align}
Finally, following the definition of $I(\epsilon_1,\epsilon_2,\beta_1,\beta_2,\lambda)$ in \eqref{eq:integral}, we can simplify \eqref{eq:app_uncon_PDF_DPSK3} into \eqref{eq:PDF_X0}.

The derivation of the PDF of $\bm{Y}_0$ follows similar lines as the derivation of $f_{\bm{X}_0}(\bm{x})$, which is summarized in two main steps. In the first step, the conditional PDF of $\bm{Y}_0$ given $|X_2|^2=z$ is obtained as
\begin{align}
  f_{\bm{Y}_0\big||X_2|^2=z}(\bm{y})&= \frac{\e^{-\frac{|y_p|^2}{(\Omega_1+\sigma_1^2)z+\sigma_2^2}}}{\pi[(\Omega_1+\sigma_1^2)z+\sigma_2^2]}\prod_{\substack{i=1\\ i\neq p}}^M\frac{\e^{-\frac{|y_i|^2}{\sigma_1^2z+\sigma_2^2}}}{\pi(\sigma_1^2z+\sigma_2^2)} \nonumber\\
  &=\frac{\e^{-\big(\frac{\|\bm{y}\|^2-|y_p|^2}{\sigma_1^2z+\sigma_2^2}+\frac{|y_p|^2}{(\Omega_1+\sigma_1^2)z+\sigma_2^2}\big)}}{\pi^M (\sigma_1^2z+\sigma_2^2)^{M-1} [(\Omega_1+\sigma_1^2)z+\sigma_2^2]}.\label{eq:app_cond_PDF_Y0_1}
\end{align}
In the second step, the unconditional PDF of $\bm{Y}_0$  is obtained as
\begin{align}
  f_{\bm{Y}_0}(\bm{y})
&=\int_0^\infty f_{\bm{Y}_0\big||X_2|^2=z}(\bm{y})f_{|X_2|^2}(z)\diff z \nonumber\\
  &=\frac{1}{(\pi\sigma_2^2)^M}\int_0^\infty \frac{\exp\Big[-\Big(z+\frac{\|\bm{y}\|^2-|y_p|^2}{\Omega_2\sigma_1^2z+\sigma_2^2}+\frac{|y_p|^2}{\Omega_2(\Omega_1+\sigma_1^2)z+
  \sigma_2^2}\Big)\Big]}{\Big(1+\frac{\Omega_2\sigma_1^2}{\sigma_2^2}z\Big)^{M-1}\Big[1+\frac{\Omega_2(\Omega_1+\sigma_1^2)}{\sigma_2^2}z\Big]}\diff z, \label{eq:app_uncon_PDF_FSK}
\end{align}
which simplifies to \eqref{eq:PDF_Y0} following the definition of $I(\epsilon_1,\epsilon_2,\beta_1,\beta_2,\lambda)$ in \eqref{eq:integral}.

\renewcommand{\thesection}{\Alph{section}}
\renewcommand{\theequation}{\Alph{section}.\arabic{equation}}
\setcounter{equation}{0}
\section{Proof of Theorem \ref{thm:Exact_MLD}}\label{app:Exact_MLD}
In this proof, we will start with the derivation of the MLD for $M$-DPSK. Then the MLD for $M$-FSK will be derived in a similar procedure.

Recall the unified signal model in \eqref{eq:DPSK_unified} for both PS-NcAF and TS-NcAF employing $M$-DPSK. It is not hard to see that $\bm{y}_{0d}|_m\sim\mathcal{CN}(\bm{0},\bm{\Sigma}_{0d})$, where
\begin{align}
  \bm{\Sigma}_{0d}=\sigma_{0d}^2\left[
                             \begin{array}{cc}
                               1+\gamma_{0d} & \gamma_{0d}\e^{-j2\pi m/M} \\
                               \gamma_{0d}\e^{j2\pi m/M} & 1+\gamma_{0d} \\
                             \end{array}
                           \right].
\end{align}
Thus, the conditional PDF of $\bm{y}_{0d}$ given $m$ can be expressed as
  \begin{align}
    f(\bm{y}_{0d}|m) = \frac{\exp\left(-\frac{(1+\gamma_{0d})\|\bm{y}_{0d}\|^2}{\sigma_{0d}^2(1+2\gamma_{0d})}\right)}{(\pi\sigma_{0d}^2)^2(1+2\gamma_{0d})}\exp\bigg\{
     \frac{2\gamma_{0d}}{1+2\gamma_{0d}}\frac{\Re\{y_{0d}(l-1)y_{0d}^*(l)\e^{j2\pi m/M}\}}{\sigma_{0d}^2}\bigg\}. \label{eq:app_LF_0d_PSK}
 \end{align}
Using some algebraic manipulations, the signal model for $\bm{y}_{rd}$ in \eqref{eq:yrd_DPSK_unified} can be rewritten in the desired form as $\bm{X}_0$ in \eqref{eq:bm_X0}
\begin{align}
  \bm{y}_{rd}=\tilde{X}_1\tilde{X}_2\tilde{\bm{c}}+\tilde{X}_2\tilde{\bm{x}}_1+\tilde{\bm{x}}_2,
\end{align}
where
\begin{subequations}
\begin{align}
    \tilde{\bm{c}}&\pardef [1,c(l)]^T\\
    \tilde{X}_1&\pardef\sigma_{0r}\sqrt{\gamma_{0r}}h_{0r}s(l-1)\sim\mathcal{CN}(0,\sigma_{0r}^2\gamma_{0r})\\
    \tilde{X}_2&\pardef\sigma_{rd}\sqrt{\gamma_{rd}}h_{rd}\sim\mathcal{CN}(0,\sigma_{rd}^2\gamma_{rd})\\
    \tilde{\bm{x}}_1&\pardef\sigma_{0r}\bm{n}_{0r}\sim\mathcal{CN}(\bm{0},\sigma_{0r}^2\bm{I}_2)\\
    \tilde{\bm{x}}_2&\pardef\sigma_{rd}\bm{n}_{rd}\sim\mathcal{CN}(\bm{0},\sigma_{rd}^2\bm{I}_2).
\end{align}
\end{subequations}
Applying Lemma \ref{lemma:PDF},  the conditional PDF of $\bm{y}_{rd}$ given $m$ can be written as
  \begin{align}
    f(\bm{y}_{rd}|m) = \frac{1}{(\pi\sigma_{rd}^2)^2}I\bigg(\sigma_{0r}^2\gamma_{rd},  ~ & (1+2\gamma_{0r})\sigma_{0r}^2\gamma_{rd},
   \frac{|y_{rd}(l)-y_{rd}(l-1)\e^{j2\pi m/M}|^2}{2\sigma_{rd}^2},\nonumber\\
   &\frac{|y_{rd}(l)+y_{rd}(l-1)\e^{j2\pi m/M}|^2}{2\sigma_{rd}^2},1\bigg). \label{eq:app_LF_rd_PSK}
 \end{align}
Following the definition of the MLD, we have
\begin{align}
  \hat{m} & = \arg\max_{m=0,\cdots,M-1} f(\bm{y}_{0d},\{\bm{y}_{rd}\}_{r=1}^K|m) \\
          & = \arg\max_{m=0,\cdots,M-1} \left \{\ln f(\bm{y}_{0d}|m)+\sum_{r=1}^K \ln f(\bm{y}_{rd}|m)\right\},\label{eq:app_MLD_DPSK1}
\end{align}
where $f(\bm{y}_{0d},\{\bm{y}_{rd}\}_{r=1}^K|m)$ is the likelihood function when $m$ is transmitted, and \eqref{eq:app_MLD_DPSK1} follows by the independence among all different signal branches. Substituting \eqref{eq:app_LF_0d_PSK} and \eqref{eq:app_LF_rd_PSK} into \eqref{eq:app_MLD_DPSK1}  yields the MLD in \eqref{eq:MLD_DPSK} for $M$-DPSK.

The MLD for $M$-FSK can be derived with similar steps as in the derivation of the MLD for $M$-DPSK. Specifically, we first obtain the PDF of $\by_{0d}$ in \eqref{eq:y0d_FSK_unified}, conditioned on the transmitted message $m$, as
\begin{align}
  f(\by_{0d}|m)=\frac{\e^{-\frac{\|\by_{0d}\|^2}{\sigma_{0d}^2}}}{(\pi\sigma_{0d}^2)^M(1+\gamma_{0d})}\e^{\frac{\gamma_{0d}}{1+\gamma_{0d}}
  \frac{|\mathrm{y}_{0d}(m+1)|^2}{\sigma_{0d}^2}}.\label{eq:app_cond_LF_0d_FSK}
\end{align}
Then, $\by_{rd}$ in \eqref{eq:yrd_FSK_unified} can be rewritten in the desired form as $\bm{Y}_0$ as $\by_{rd}=\breve{X}_1\breve{X}_2\bi_{m+1}+\breve{X}_2\breve{\bm{y}}_1+\breve{\bm{y}}_2$, where $\breve{X}_1\pardef \sigma_{0r}\sqrt{\gamma_{0r}}h_{0r}\sim\mathcal{CN}(0,\sigma_{0r}^2\gamma_{0r})$, $\breve{X}_2\pardef \sigma_{rd}\sqrt{\gamma_{rd}}h_{rd}\sim\mathcal{CN}(0,\sigma_{rd}^2\gamma_{rd})$, $\breve{\bm{y}}_1 \pardef\sigma_{0r}\bn_{0r}\sim\mathcal{CN}(\bm{0},\sigma_{0r}^2\bm{I}_M)$, and $\breve{\bm{y}}_2 \pardef\sigma_{rd}\bn_{rd}\sim\mathcal{CN}(\bm{0},\sigma_{rd}^2\bm{I}_M)$. Then, applying Lemma \ref{lemma:PDF},  the conditional PDF of $\by_{rd}$ given $m$ can be expressed as
  \begin{align*}
    f(\by_{rd}|m) = \frac{1}{(\pi\sigma_{rd}^2)^M}I\bigg(\sigma_{0r}^2\gamma_{rd},(1+\gamma_{0r})\sigma_{0r}^2\gamma_{rd}, \frac{\|\by_{rd}\|^2-|\mathrm{y}_{rd}(m+1)|^2}{\sigma_{rd}^2}, \frac{|\mathrm{y}_{rd}(m+1)|^2}{\sigma_{rd}^2},M-1\bigg). 
 \end{align*}
Finally, following the definition of MLD, we can show that the MLD for $M$-FSK is given by \eqref{eq:MLD_FSK}.

\renewcommand{\thesection}{\Alph{section}}
\renewcommand{\theequation}{\Alph{section}.\arabic{equation}}
\setcounter{equation}{0}
\section{Proof of Theorem \ref{th:Approx_MLD}}\label{app:Approx_MLD}
The main difficulty involved in the MLD is the integral computation of $I(\epsilon_1,\epsilon_2,\beta_1,\beta_2,\lambda)$ in \eqref{eq:integral}. Thus, we will develop a tight closed-form approximation for this integral. To that end, we first transform the original integral into a desired form which is easy to approximate. Specifically, we carry out the following algebraic operations:
\begin{align}
  I(\epsilon_1,\epsilon_2,\beta_1,\beta_2,\lambda)&=\int_0^\infty\frac{\e^{-\big(x+\frac{\beta_1}{1+\epsilon_1 x}+\frac{\beta_2}{1+\epsilon_2 x}\big)}}{(1+\epsilon_2x)^\lambda(1+\epsilon_2x)}\diff x \label{eq:app_integral1} \\
  &=\int_0^1\frac{\e^{-\big(\frac{\beta_1}{1-\epsilon_1\ln z}+\frac{\beta_2}{1-\epsilon_2\ln z}\big)}}{(1-\epsilon_1\ln z)^\lambda(1-\epsilon_2\ln z)}\diff z\label{eq:app_integral2}\\
  &=\int_0^1 \psi(z) \diff z,\label{eq:app_integral3}
\end{align}
where \eqref{eq:app_integral2} follows by the change of variable $\e^{-x}\rightarrow z$ and $\psi(z)$ is defined in \eqref{eq:Int_approx_subfunc}. The integrand $\psi(z)$ is a well-behaved function: first, it is a bounded positive function with $0<\psi(z)<1$; second, it is a smooth function with continuous derivatives with respect to $z$ of any order. Furthermore, the integral limits are finite. These elegant properties enable very accurate approximation of the integral using the GL quadrature with a finite order. Indeed, the choice of the approximation order must always compromise between the accuracy and computational cost  \cite{P.J.Davis1984book}. The higher order, the more accurate the GL quadrature is. On the other hand, however, higher order requires the computation of more terms which involves higher computational cost. Considering both the accuracy and computational cost, we choose the fifth order GL quadrature for our purpose \cite{P.J.Davis1984book}, which results in $I(\epsilon_1,\epsilon_2,\beta_1,\beta_2,\lambda)\approx \tilde{I}(\epsilon_1,\epsilon_2,\beta_1,\beta_2,\lambda)$, where $\tilde{I}(\cdot,\cdot,\cdot,\cdot,\cdot)$ is given in closed-form in \eqref{eq:integral_app}. Finally, using $\tilde{I}(\cdot,\cdot,\cdot,\cdot,\cdot)$ in place of $I(\cdot,\cdot,\cdot,\cdot,\cdot)$ in the MLDs yields the corresponding GLDs.\footnote{Simulation results in Section \ref{sec:numerical} demonstrate that the performance of GLDs are almost the same as that of the MLDs, indicating that the fifth order GL approximation adopted here is highly accurate.}

\end{document}